# Coupled spin-1/2 ladders as microscopic models for non-Abelian chiral spin liquids

Po-Hao Huang,[1] Jyong-Hao Chen,[2] Adrian E. Feiguin,[3] Claudio Chamon,[1] and Christopher Mudry[2]

[1] *Department of Physics, Boston University, Boston, MA, 02215, USA*
[2] *Condensed Matter Theory Group, Paul Scherrer Institute, CH-5232 Villigen PSI, Switzerland*
[3] *Department of Physics, Northeastern University, Boston, Massachusetts 02115, USA*

We construct a two-dimensional (2D) lattice model that is argued to realize a gapped chiral spin liquid with (Ising) non-Abelian topological order. The building blocks are spin-1/2 two-leg ladders with $SU(2)$-symmetric spin-spin interactions. The two-leg ladders are then arranged on rows and coupled through $SU(2)$-symmetric interactions between consecutive ladders. The intra-ladder interactions are tuned so as to realize the $c = 1/2$ Ising conformal field theory, a fact that we establish numerically via Density Matrix Renormalization Group (DMRG) studies. Time-reversal breaking inter-ladder interactions are tuned so as to open a bulk gap in the 2D lattice system. This 2D system supports gapless chiral edge modes with Ising non-Abelian excitations but no charge excitations, in contrast to the Pfaffian non-Abelian fractional quantum Hall state.

That point particles may obey non-Abelian braiding statistics in (2+1)-dimensional spacetime has been known in quantum-field theory since the 1980's [1–5]. Moore and Read showed in 1991 that certain Pfaffian wave functions support quasi-particles with non-Abelian braiding statistics [6]. This discovery opened the possibility that non-Abelian braiding statistics could be found in certain fractional quantum Hall plateaus [6–9].

A second physical platform to realize braiding statistics that is neither bosonic nor fermionic is provided by quasi-two-dimensional quantum spin magnets with a gapped chiral spin-liquid ground state [10, 11]. Quasi-two-dimensional arrays of quantum spin chains also have the potential for realizing gapped spin liquid ground states with quasi-particles obeying Abelian or non-Abelian braiding statistics [12–15].

In this paper, we construct a two-dimensional (2D) lattice model, depicted in Fig. 1, that is argued to realize a non-Abelian chiral spin liquid. This 2D model consists of an array of coupled one-dimensional (1D) two-leg quantum spin-1/2 ladders. The inter-ladder coupling leads to a bulk gap, while gapless modes remain at the boundaries. The chiral edge states correspond to the Ising conformal field theory (CFT) with central charge $c = 1/2$, similarly to the Moore-Read Pfaffian state. However, in contrast to the Pfaffian quantum Hall state, there is no additional $c = 1$ chiral bosonic charge-carrying edge mode. By the bulk-edge correspondence, the bulk of the coupled spin-ladder model is a gapped chiral spin liquid supporting Ising non-Abelian topological order [16, 17].

To obtain this result, we argue that the aforementioned lattice model is a regularization of one of the interacting quantum-field theories presented in Ref. 14, one that supports chiral non-Abelian topological order. We start from coupled two-leg ladders (called bundles in Ref. 14), on which sites quantum spin-1/2 degrees of freedom are localized. Two ingredients are needed. First, the interactions within the two-leg ladders should be fine-tuned so as to realize the Ising universality class in (1+1)-dimensional spacetime, the Ising criticality in short. Second, the interactions between the two-leg ladders (the bundles) should be dominated by *strong* current-current interactions. Alternatively, the interactions between the ladders (bundles) could be weak when mediated by Kondo-like quantum spin-1/2 degrees of freedom, as shown by Lecheminant and Tsvelik in Ref. 15. If so, the results of Ref. 14 imply that the 1D array of coupled two-leg ladders is a lattice regularization of a chiral spin liquid supporting Ising non-Abelian topological order. We are going to detail how we achieve Ising criticality in a single two-leg ladder, and how these fine-tuned two-leg ladders are coupled so as to stabilize 2D Ising non-Abelian topological order.

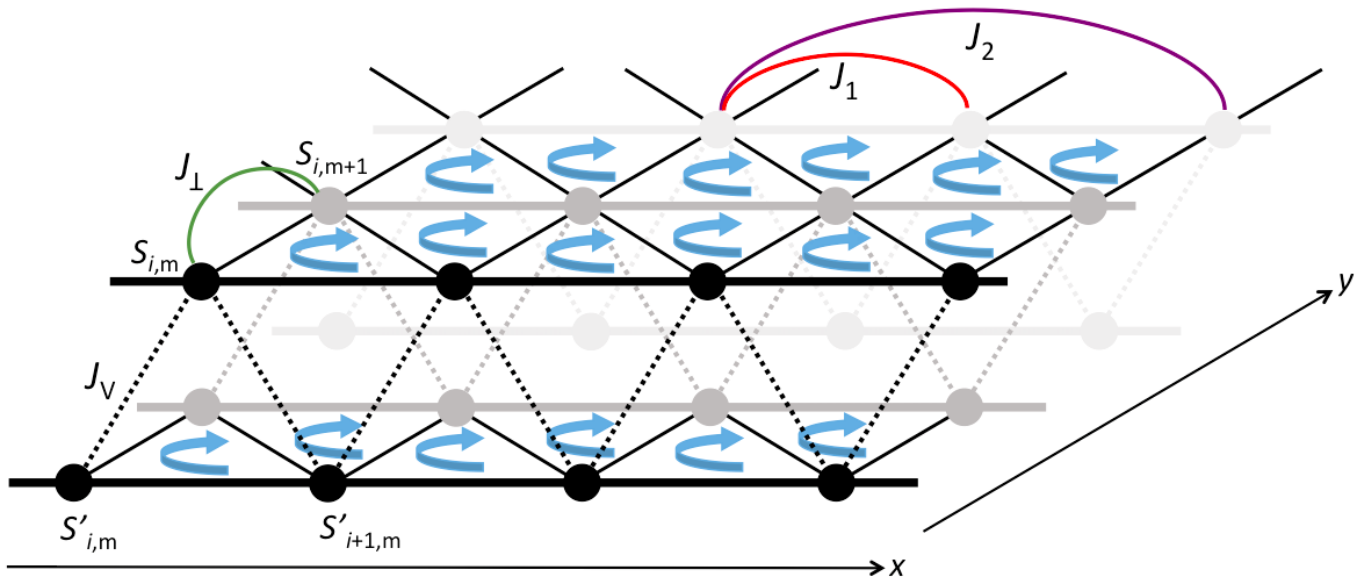

FIG. 1. (Color online) Strongly coupled spin-1/2 two-leg ladders that realize the Ising topological order in two-dimensional space. The intra-ladder couplings $J_1$, $J_2$, and $J_\vee$ are defined in Eq. (1). The inter-ladder couplings $J_\perp$ (represented by the green bond) and $\chi_\perp$ (represented by the blue arrows) are defined in Eq. (6).

**Ising criticality in a ladder** – To realize the Ising criticality, we assume that the intra-ladder spin-1/2 interactions allow to interpolate between two distinct dimerized ground states. This can be achieved by positing the following quantum spin-1/2 Hamiltonian on a two-leg ladder

$$\widehat{H}_{\rm ladder} := \widehat{H}_{\rm leg} + \widehat{H}'_{\rm leg} + \widehat{H}_{\rm zig-zag}. \qquad (1a)$$

The first leg of the ladder hosts the quantum spin-1/2 operators $\widehat{\boldsymbol{S}}_i$ on every site $i = 1, \cdots, N$, where any two consecutive sites is displaced by the lattice spacing $\mathfrak{a}$. Similarly, the second leg of the ladder hosts spin-1/2 operators $\widehat{\boldsymbol{S}}'_{i'}$ on every site $i' = 1, \cdots, N$. Hamiltonians $\widehat{H}_{\rm leg}$ and $\widehat{H}'_{\rm leg}$ are the antiferromagnetic $J_1$-$J_2$ one-dimensional Heisenberg model, i.e.,

$$\widehat{H}_{\rm leg} := \sum_{i=1}^{N} \left( J_1\, \widehat{\boldsymbol{S}}_i \cdot \widehat{\boldsymbol{S}}_{i+1} + J_2\, \widehat{\boldsymbol{S}}_i \cdot \widehat{\boldsymbol{S}}_{i+2} \right) \tag{1b}$$

with $J_1, J_2 \geq 0$ and $\widehat{H}'_{\rm leg}$ obtained from $\widehat{H}_{\rm leg}$ with the substitution $\widehat{\boldsymbol{S}}_i \to \widehat{\boldsymbol{S}}'_{i'}$. The spin-1/2 operators on the two legs also interact through a $SU(2)$-symmetric antiferromagnetic Heisenberg exchange interaction, which we choose to be

$$\widehat{H}_{\rm zig-zag} := J_{\vee} \sum_{i,i'=1}^{N} \left( \delta_{i',i} + \delta_{i',i+1} \right) \widehat{\boldsymbol{S}}_i \cdot \widehat{\boldsymbol{S}}'_{i'} \tag{1c}$$

with $J_{\vee} \geq 0$. The coordination number in $\widehat{H}_{\rm zig-zag}$ is two, not one as would be the case for the standard rung antiferromagnetic Heisenberg exchange interaction.

Hamiltonian (1) is invariant under a global $SU(2)$ rotation of all spins, the interchange of the upper and lower legs, and the mirror symmetries centered about a site of one leg and the middle of the bond of the other leg. Hamiltonian (1) simplifies in two limits, namely when $J_{\vee} = 0$ or when $J_2 = 0$.

When $J_{\vee} = 0$, Hamiltonian (1) is the sum of two independent $J_1$-$J_2$ antiferromagnetic Heisenberg chains. It is gapless when $J_2/J_1 \leq (J_2/J_1)_{\rm c} \approx 0.24$ and gapful otherwise [18, 19]. In the gapped phase, the ground state manifold is four-fold degenerate as the translation symmetry along each leg is spontaneously broken by one of two possible (leg) dimerized ground states when periodic boundary conditions are imposed (by identifying site $N+1$ with site 1). In particular, at the Majumdar-Ghosh point $J_2/J_1 = 1/2$ [20], the ground state is a linear combination of the four possible direct products of all singlet states of two spin-1/2 degrees of freedom on every other bond along the upper or lower legs.

When $J_2 = 0$, Hamiltonian (1) is the $J_{\vee}$-$J_1$ antiferromagnetic Heisenberg quantum spin-1/2 zig-zag chain. (Notice that the zig-zag chain is equivalent to one of the chains discussed above upon the identification $J_1 \to J_2$ and $J_{\vee} \to J_1$.) The zig-zag chain is gapless when $J_1/J_{\vee} \leq (J_1/J_{\vee})_{\rm c} \approx 0.24$ and gapful otherwise [18]. In the gapped phase, the ground state manifold is two-fold degenerate as the translation symmetry along the chain is spontaneously broken by one of two possible (zig-zag) dimerized ground states when periodic boundary conditions are imposed. Again, at the Majumdar-Ghosh point $J_1/J_{\vee} = 1/2$ [20], the ground state is a linear combination of the two possible direct products of all singlet states of two spin-1/2 degrees of freedom on every other bond along the zig-zag chain. (Henceforth, we shall measure all energies in units of $J_1$, i.e., $J_1 \equiv 1$.)

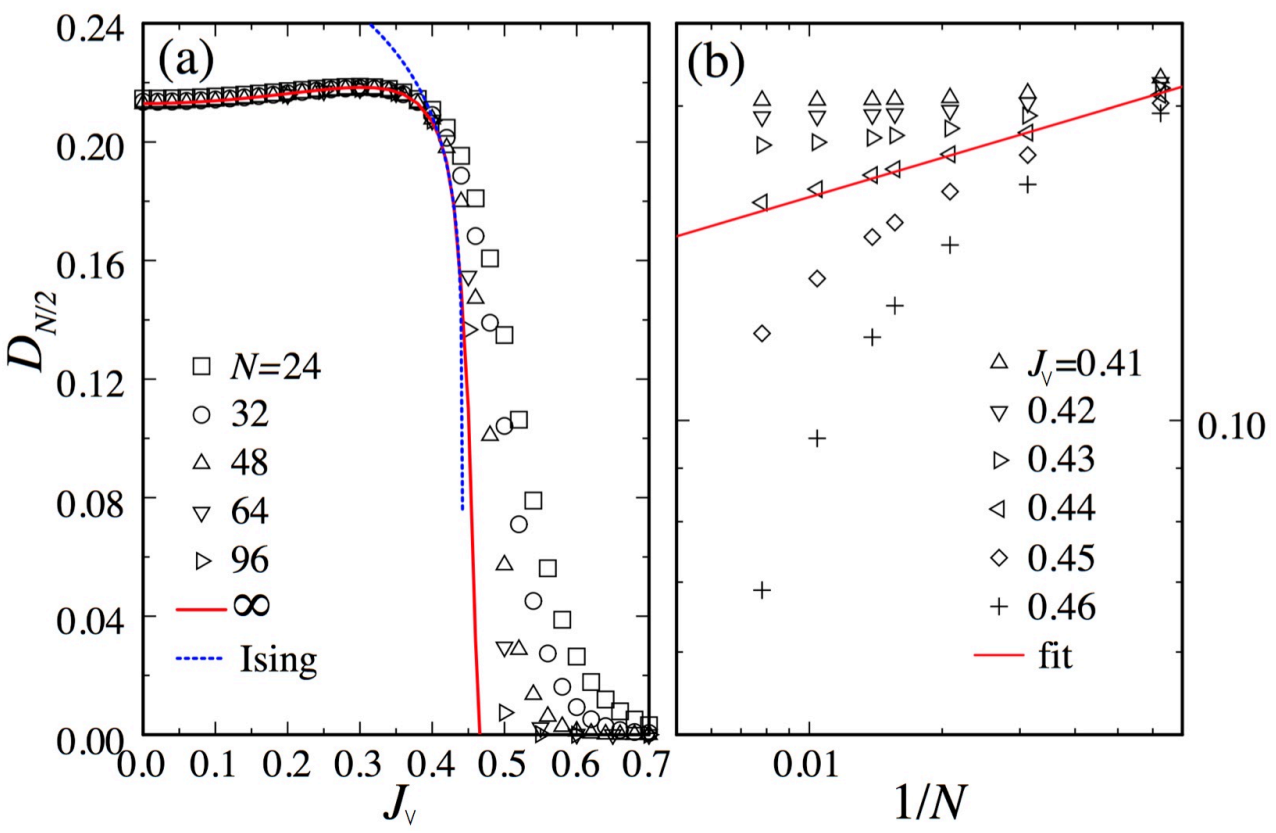

FIG. 2. (Color online) (a) Leg-dimer order parameter at the center of the ladder, $D_{N/2}$, across the transition for fixed value of $J_2 = 0.44$ and different system sizes. The extrapolation to the thermodynamic limit is obtained with a second order polynomial in $1/N$, while the dashed curve is a fit to the Ising scaling law with a transition at $J_{\vee} = 0.442$. (b) Scaling of $D_{N/2}$ with $1/N$ for different values of $J_{\vee}$ around the critical point. The scaling at the transition ($J_{\vee} = 0.44$) is well described by an exponent 1/8 corresponding to the Ising universality class.

We are now going to present numerical evidence according to which two distinct dimerized phases that are adiabatically connected to the two Majumdar-Ghosh points $(J_2, J_{\vee}) = (0, 2)$ and $(1/2, 0)$, respectively, are separated by a phase boundary that realizes the Ising criticality. This is to say that it is possible to connect the Majumdar-Ghosh points $(J_2, J_{\vee}) = (0, 2)$ and $(1/2, 0)$ by a one dimensional path in parameter space along which the spectral gap and order parameters vanish at an isolated quantum critical point. We note that a transition between distinct dimerized phases may be a general route to achieving an Ising critical theory. Vekua and Honecker in Ref. 21 and Lavarélo et al. in Ref. 22 have also shown numerically that two other families of ladders for quantum spin-1/2 display a quantum critical point in the Ising universality class that separates two distinct gapped dimer phases [23]. We note as well that quantum spin-1 chains can also show a quantum critical point in the Ising universality class separating gapped phases that are not related by a loss of symmetry [24].

In order to determine the phase diagram and the nature of the quantum transitions between the different phases at zero temperature, we resort to the density matrix renormalization group method (DMRG) [25, 26]. We simulate the two-leg ladder (1) with open boundary conditions using up to 2000 DMRG states for the largest systems considered ($N = 768$), which guarantees an ac-curacy of 9 significant digits in the energy, and 6 significant digits in the entanglement entropy. We focus on the transition line that separates the leg-dimer and zig-zag-dimer phases. As in Ref. 22, we plot the ground-state expectation value for the leg-dimer order parameters in Fig. 2(a) as a function of $J_\vee$ for different system sizes, together with an extrapolation to the thermodynamic limit, which allows us to locate the transition at the point $J_\vee \approx J_2 \approx 0.44$ [27]. Panel (b) shows the anomalous scaling exponent of the leg-dimer order parameter as a function of $1/N$. It can be approximated by an exponent of 1/8 precisely at $J_2 = 0.44$, indicating the Ising nature of the transition.

Further evidence for the nature of the transition is found through finite-size scaling of the energy spectra and the entanglement entropy.

The finite size spectrum for the 2D Ising CFT depends on the boundary conditions [28]. For open boundary condition, CFT predicts the spectrum

$$E_n(N) = \varepsilon_0\, N + \varepsilon_1 + \frac{\pi v}{N}\left(-\frac{1}{48} + x_n\right) + \mathcal{O}\left(\frac{1}{N^2}\right), \quad (2)$$

where $\varepsilon_0$ and $\varepsilon_1$ are non-universal constants, while $x_n$ is the anomalous scaling dimension of the operator corresponding to the state labeled by the integer $n$. The value of $x_n$ is sensitive to the limit $N \to \infty$ being taken with even or odd values of $N$. Which set (conformal tower) of anomalous scaling dimensions enters on the right-hand side of Eq. (2) depends on the parity in the number of spins per chain in the two-leg ladder. For an even number $N$ of spins per chain, the conformal tower starts from the identity operator, i.e., $x_0 = 0$. There follows the anomalous scaling exponents $x_1 = 2$, $x_2 = 3$, $x_3 = x_4 = 4$ and so on. For an odd number of spins per chain $N$, the conformal tower starts from the energy operator $\epsilon$ with the anomalous scaling exponent $x_0 = 1/2$, followed by the exponents $x_1 = 3/2$, $x_2 = 5/2$, $x_3 = 7/2$, $x_4 = 9/2$ and so on. Our DMRG results are summarized in Figs. 3(a-d). For any $n = 0, 1, 2, 3, 4$, analyzing the leading linear dependence and the axis intercept of $E_n(N)/N$ *vs* $1/N$ determines the numerical values of $\varepsilon_0$ and $\varepsilon_1$. The value of $v$ is obtained from averaging the slope of $[E_n(N) - \varepsilon_0\, N - \varepsilon_1]/N$ as a function of $1/N^2$ in Eq. (2) for $n = 0, 1, 2, 3, 4$ assuming that $x_n$ is governed by the Ising universality class. The consistency of this assumption is then verified by fitting $x_n$ from the slope of $[E_n(N) - \varepsilon_0\, N - \varepsilon_1]/N$ as a function of $1/N^2$ with $v$ given as above. Alternatively, $x_n$ can be fitted from the slope of $E_n(N) - E_0(N)$ as a function of $1/N$ with $v$ given as above. For even and odd $N$, the values $x_0 = 0.000(4)$ $x_1 = 2.0(1)$, $x_2 = 2.9(5)$, $x_3 = 3.9(2)$, $x_4 = 3.9(9)$ and $x_0 = 0.5(2)$ $x_1 = 1.5(1)$, $x_2 = 2.5(0)$, $x_3 = 3.3(9)$ follow from these fittings, respectively. They agree with the Ising universality class within the error bars.

The entanglement entropy computed with DMRG also agrees with that of the Ising universality class. If we cut open the two-leg ladder of length $N$ along a rung into one block of size $x$, the entanglement entropy $S(x, N)$ scales with $x$ and $N$ like [22, 29–33]

$$S(x, N) = \frac{c}{6} \ln\left(\frac{N+1}{\pi} \sin\frac{\pi\, x}{N+1}\right) + A\,\langle \widehat{\boldsymbol{S}}_x \cdot \widehat{\boldsymbol{S}}_{x+1}\rangle + B, \quad (3)$$

where the number $c = 1/2$ is the Ising central charge, while $A$ and $B$ are non-universal constants. The entanglement entropy $S(x, N)$ and the spin-spin correlation $\langle \widehat{\boldsymbol{S}}_x \cdot \widehat{\boldsymbol{S}}_{x+1}\rangle$ are computed by DMRG and fitted according to the scaling law (3) as summarized in Figs. 3(e-f). The best fit for $c$ is very close to one half irrespective of whether $x$ is varied holding $N$ fixed or choosing $x = N/2$.

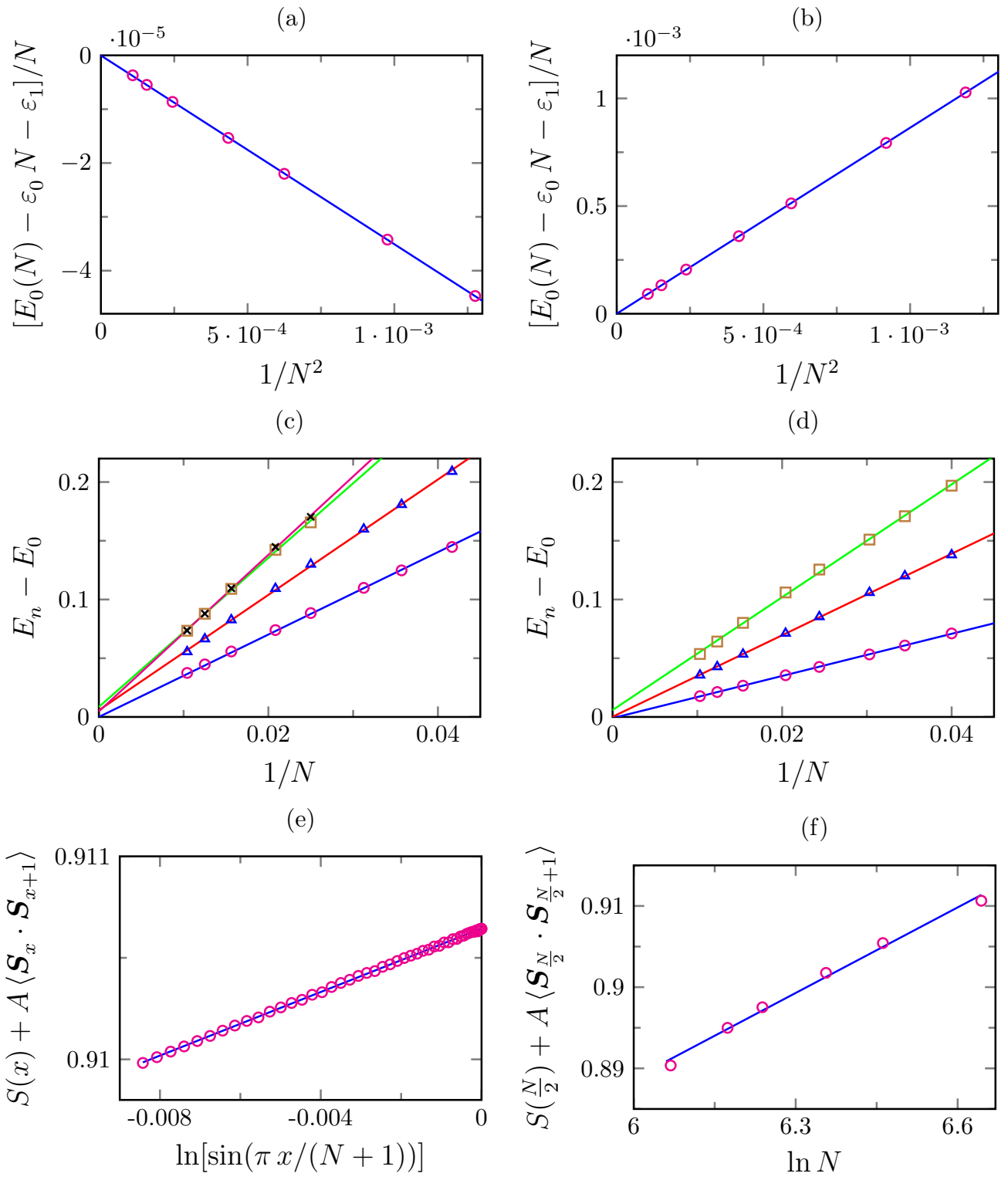

FIG. 3. (Color online) The function $[E_0(N) - \varepsilon_0\, N - \varepsilon_1]/N$ is plotted as a function of $1/N^2$ with (a) $N$ even and (b) $N$ odd. The non-universal constants $\varepsilon_0 = 0.7771(2)$ and $\varepsilon_1 = 0.040(6)$ follow from a linear dependence of $E_n(N)/N$ on $1/N$ intercepting the origin in the thermodynamic limit $N \to \infty$. The slopes in (a) and (b) give $\pi\, v\, [x_0 - (1/48)]$ for $N$ even and odd, respectively. The function $E_n(N) - E_0(N)$ is plotted as a function of $1/N$ with (c) $N$ even and (d) $N$ odd for $n = 1, 2, 3, 4$ and $n = 1, 2, 3$, respectively. The slopes in (c) and (d) give $\pi\, v\, (x_n - x_0)$ when the limit $N \to \infty$ is taken with $N$ even and odd, respectively. The slopes from the plots of Eq. (3) as a function of (e) $\sin\big(\pi\, x/(N+1)\big)$ with $N = 768$ fixed and (f) $\ln N$ with $x$ fixed yield $c = 0.4(7)$ and $c = 0.4(9)$, respectively.

We close this discussion of a single two-leg ladder by providing a field-theory description of the Ising critical-ity. The continuum field theory shines light into why Ising criticality emerges, and into how to couple the two-leg ladders together so as to build the 2D bulk-gapped topological phase in the second step of our construction. The quantum fields that encode the low energy degrees of freedom around the Ising critical point follow from the identifications (see the Supplementary Material and Ref. 34)

$$\widehat{\boldsymbol{S}}_i \to \widehat{\boldsymbol{J}}_L(x) + \widehat{\boldsymbol{J}}_R(x) + (-1)^i\, \widehat{\boldsymbol{n}}(x), \tag{4a}$$

$$\widehat{\boldsymbol{S}}'_{i'} \to \widehat{\boldsymbol{J}}'_L(x) + \widehat{\boldsymbol{J}}'_R(x) + (-1)^{i'}\, \widehat{\boldsymbol{n}}'(x). \tag{4b}$$

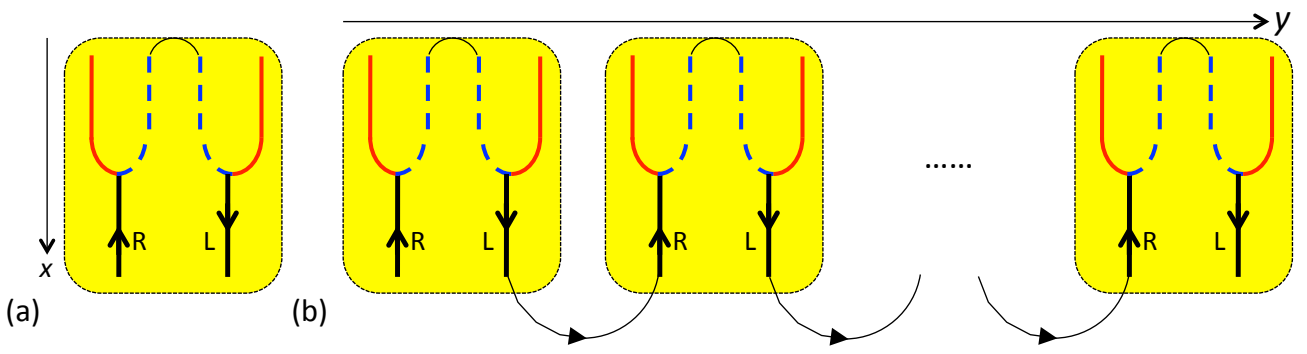

FIG. 4. (Color online) (a) Coset representation of the critical point of the two-leg ladder in the Ising universality class. (b) Coset representation of the strong current-current interactions that stabilize the chiral spin liquid phase with Ising topological order. A microscopic regularization of these strong current-current interactions is encoded by the inter-ladder couplings $J_\perp$ and $\chi_\perp$ defined in Eq. (6).

The modes that vary slowly on the scale of the lattice spacing $\mathfrak{a}$ are the non-Abelian chiral currents $\widehat{\boldsymbol{J}}_M(x)$ and $\widehat{\boldsymbol{J}}'_M(x)$ with $M = L, R$ on the upper and lower legs, respectively. Their scaling dimension is 1 when $J_2 = J_\vee = 0$. The quantum fields $\widehat{\boldsymbol{n}}(x)$ and $\widehat{\boldsymbol{n}}'(x)$ represent the staggered magnetizations on their respective legs. Their scaling dimension is 1/2 when $J_2 = J_\vee = 0$. In the absence of the microscopic couplings $J_2$ and $J_\vee$, each chain can be separately described using the $\widehat{su}(2)_1$ affine Lie algebra satisfied by the chiral currents. Together, the two sets of currents also satisfy a $\widehat{su}(2)_1 \oplus \widehat{su}(2)_1$ affine Lie algebra (with central charge $c = 2$). Once the microscopic couplings $J_2$ and $J_\vee$ are turned on, a number of macroscopic interactions appear, including the marginally relevant current-current interaction $(\widehat{\boldsymbol{J}}_L + \widehat{\boldsymbol{J}}'_L) \cdot (\widehat{\boldsymbol{J}}_R + \widehat{\boldsymbol{J}}'_R)$. The chiral sums $\widehat{\boldsymbol{J}}_M + \widehat{\boldsymbol{J}}'_M$, $M = L, R$, of the currents on both chains satisfy themselves an affine sub-algebra $\widehat{su}(2)_2$ (with central charge $c = 3/2$). At strong coupling, the added interactions gap the $\widehat{su}(2)_2$ piece, leaving behind the coset theory $[\widehat{su}(2)_1 \oplus \widehat{su}(2)_1]/\widehat{su}(2)_2$, which is precisely the Ising critical theory (with central charge $c = 1/2$). While the marginal twist term $\boldsymbol{n} \cdot \partial_x \boldsymbol{n}'$ is allowed by symmetry in the continuum description of the two-leg ladder [35], our DMRG results support the case that, by properly selecting the microscopic couplings, the total current-current interactions can dominate the renormalization-group (RG) flow to strong couplings and gap the corresponding sub-algebra.

A useful pictorial rendition of this mechanism is the following. The two-leg ladder is represented by a colored square box in Fig. 4(a). The chiral critical modes generating the affine Lie algebra $\widehat{su}(2)_1 \oplus \widehat{su}(2)_1$ are represented by two lines with opposite arrow in Fig. 4(a). The forking of either one of the directed lines represents the fact that the affine Lie algebra contains the diagonal affine subalgebra $\widehat{su}(2)_{1+1}$ (dashed blue tine of the fork) and the coset $\widehat{su}(2)_1 \oplus \widehat{su}(2)_1/\widehat{su}(2)_{1+1}$ (red tine). The marginally relevant current-current perturbation $(\widehat{\boldsymbol{J}}_L + \widehat{\boldsymbol{J}}'_L) \cdot (\widehat{\boldsymbol{J}}_R + \widehat{\boldsymbol{J}}'_R)$ is represented by an arc that connects the lines associated with the $\widehat{su}(2)_{1+1}$ subalgebra. This coupling gaps the modes associated with this subalgebra without affecting the modes associated with the coset theory. We are thus left with a gapless Ising critical theory.

**Ising non-Abelian topological order** – Equipped with this pictorial representation, we consider next an array of two-leg ladders labeled by the index $\mathtt{m} = 1, \cdots, n$ in Fig. 4(b), each of which is fine-tuned to the Ising quantum critical point. Next, we present a mechanism to gap the bulk modes, and leave behind only the Ising critical theories at the left-most and right-most bundles, i.e., at the edges. This cannot be achieved by simply coupling the Ising modes with opposite chirality across any two consecutive two-leg ladders, depicted as neighboring colored boxes in Fig. 4(b). The reason is that the Ising degrees of freedom are fractionalized, and one is only allowed to write microscopic couplings between unfractionalized degrees of freedom. There is no physical current operator associated with the coset. The mechanism to circumvent this problem was presented in Ref. 14. One couples the chiral currents associated with the original $\widehat{su}(2)_1 \oplus \widehat{su}(2)_1$ algebra with the same end result of gapping the bulk and leaving the edge states.

To gap the bulk, any two consecutive two-leg ladders are coupled by the marginally relevant $\widehat{su}(2)_1 \oplus \widehat{su}(2)_1$ current-current interactions

$$\widehat{\mathcal{H}}_{jj} := g_{jj} \sum_{\mathtt{m}=1}^{n-1} \left( \widehat{\boldsymbol{J}}_{L,\mathtt{m}} \cdot \widehat{\boldsymbol{J}}_{R,\mathtt{m}+1} + \widehat{\boldsymbol{J}}'_{L,\mathtt{m}} \cdot \widehat{\boldsymbol{J}}'_{R,\mathtt{m}+1} \right). \tag{5}$$

These couplings are represented by the directed arcs in Fig. 4(b). The arrows on the arcs are needed because this choice of current-current interaction breaks time-reversal symmetry. By inspection of Fig. 4(b), the array of two-leg ladders is fully gapped if periodic boundary conditions are imposed on the label $\mathtt{m}$, whereas gapless chiral edge states from the Ising universality class survive in the vicinity of the first and last two-leg ladders when open boundary conditions hold. Hereto, we want a lattice regularization of the quantum-field theory represented by Fig. 4(b). It is depicted in Fig. 1.

Microscopic interactions between neighboring two-leg ladders that break time-reversal symmetry and generate the desired current-current interactions (5) are obtained

from the interactions

$$\widehat{H}_{\text{inter-ladder}} := \widehat{H}_{\triangle} + \widehat{H}'_{\triangle} + \widehat{H}_{\square} + \widehat{H}'_{\square} \,, \tag{6a}$$

where

$$\begin{aligned}\widehat{H}_{\triangle} := \frac{\chi_\perp}{2} \sum_{i=1}^{N}\sum_{\mathtt{m}=1}^{n-1} \Big[ & \widehat{\boldsymbol{S}}_{i,\mathtt{m}+1} \cdot \Big(\widehat{\boldsymbol{S}}_{i+1,\mathtt{m}} \wedge \widehat{\boldsymbol{S}}_{i,\mathtt{m}}\Big) \\ & + \widehat{\boldsymbol{S}}_{i+1,\mathtt{m}} \cdot \Big(\widehat{\boldsymbol{S}}_{i,\mathtt{m}+1} \wedge \widehat{\boldsymbol{S}}_{i+1,\mathtt{m}+1}\Big) \Big]\end{aligned} \tag{6b}$$

and

$$\widehat{H}_{\square} := J_\perp \sum_{i=1}^{N}\sum_{\mathtt{m}=1}^{n-1} \Big(\widehat{\boldsymbol{S}}_{i,\mathtt{m}} \cdot \widehat{\boldsymbol{S}}_{i,\mathtt{m}+1} + \widehat{\boldsymbol{S}}_{i,\mathtt{m}+1} \cdot \widehat{\boldsymbol{S}}_{i+1,\mathtt{m}}\Big), \tag{6c}$$

with $\widehat{H}'_{\triangle}$ and $\widehat{H}'_{\square}$ following from $\widehat{H}_{\triangle}$ and $\widehat{H}_{\square}$ by the substitution $\widehat{\boldsymbol{S}}_{i,\mathtt{m}} \to \widehat{\boldsymbol{S}}'_{i',\mathtt{m}}$. The choice $(\chi_\perp/\pi) = 2J_\perp$ yields the current-current interaction Eq. (5) in the continuum limit with $g_{jj} \propto (\chi_\perp/\pi) + 2J_\perp$, as we show in the Supplementary Material, where we also argue that *any relevant bare coupling vanishes by symmetry.* Marginally relevant and relevant couplings that compete with $g_{jj}$ can still be generated at higher loop order, as shown in [36, 37]. If the coupling $g_{jj}$ is small, these competing interactions may overtake it in a weak coupling RG analysis. An alternative way to rephrase the issue is that, if the gap is only exponentially small in the bare coupling $g_{jj}$, the stability of the desired phase is still subject to the weak coupling analysis done around the fixed point defined by $J_\perp = \chi_\perp = 0$, i.e., a fixed point that is not stable against these competing relevant perturbations. However, for *strongly* coupled chains, the gap is *not* exponentially small in the bare coupling $g_{jj}$, and the addition of the very weak perturbations will not destroy the gap. Ultimately, non-perturbative techniques such as DMRG are needed to confirm that the Ising topological phase depicted in Fig. 4(b) is stable when interactions are strong in the microscopic model in Fig. 1 [38]. We also note that introducing a buffer of Kondo spin-1/2 between every neighboring two-leg ladders that mediate an indirect interaction between neighboring two-leg ladders, as was done by Lecheminant and Tsvelik in Ref. 15, also stabilizes the Ising topological phase depicted in Fig. 4(b).

In summary, field-theoretical arguments supported by DMRG show that it is possible to tune a quantum spin-1/2 two-leg ladder to Ising quantum criticality through *strong* current-current interactions. Similarly, *strong* current-current interactions between consecutive two-leg ladders are argued to stabilize a ground state supporting 2D Ising topological order.

**Acknowledgments** We benefited from useful discussions with A. M. Tsvelik and T. Neupert. This work is supported by DOE Grant DE-FG02-06ER46316 (P.-H.H. and C.C.), FN-SNF Grant 2000021 153648 (J.-H.C. and C.M.), and AEF acknowledges U.S. Department of Energy, Office of Basic Energy Sciences, for support under grant DE-SC0014407. We acknowledge the Condensed Matter Theory Visitors Program at Boston University for support.

[1] J. Fröhlich, in *Nonperturbative Quantum Field Theory*, NATO ASI Series, Vol. 185, edited by G. t Hooft, A. Jaffe, G. Mack, P. Mitter, and R. Stora (Springer US, 1988) pp. 71–100.
[2] J. Fröhlich and F. Gabbiani, Rev. Math. Phys. **02**, 251 (1990).
[3] J. Fröhlich, F. Gabbiani, and P.-A. Marchetti, in *The Algebraic Theory of Superselection Sectors. Introduction and Recent Results*, edited by D. Kastler (World Scientific, Singapore, 1990) p. 259.
[4] K. Rehren, in *The Algebraic Theory of Superselection Sectors. Introduction and Recent Results*, edited by D. Kastler (World Scientific, Singapore, 1990) p. 333.
[5] J. Fröhlich and P.A. Marchetti, Nuclear Physics B **356**, 533 (1991).
[6] G. Moore and N. Read, Nucl. Phys. **B360**, 362 (1991).
[7] X. G. Wen, Phys. Rev. Lett. **66**, 802 (1991).
[8] A. Kitaev, Annals of Physics **321**, 2 (2006).
[9] C. Nayak, S. H. Simon, A. Stern, M. Freedman, and S. Das Sarma, Rev. Mod. Phys. **80**, 1083 (2008).
[10] V. Kalmeyer and R. B. Laughlin, Phys. Rev. Lett. **59**, 2095 (1987).
[11] X. G. Wen, F. Wilczek, and A. Zee, Phys. Rev. B **39**, 11413 (1989).
[12] G. Gorohovsky, R. G. Pereira, and E. Sela, Phys. Rev. B **91**, 245139 (2015).
[13] T. Meng, T. Neupert, M. Greiter, and R. Thomale, Phys. Rev. B **91**, 241106 (2015).
[14] P.-H. Huang, J.-H. Chen, P. R. S. Gomes, T. Neupert, C. Chamon, and C. Mudry, Phys. Rev. B **93**, 205123 (2016).
[15] P. Lecheminant and A. M. Tsvelik, ArXiv e-prints (2016), arXiv:1608.05977 [cond-mat.str-el].
[16] X.-G. Wen, Int. J. Mod. Phys. B **05**, 1641 (1991).
[17] M. Oshikawa and T. Senthil, Phys. Rev. Lett. **96**, 060601 (2006).
[18] F. D. M. Haldane, Phys. Rev. B **25**, 4925 (1982).
[19] S. R. White and I. Affleck, Phys. Rev. B **54**, 9862 (1996).
[20] C. K. Majumdar and D. K. Ghosh, J. Math. Phys. **10**, 1399 (1969).
[21] T. Vekua and A. Honecker, Phys. Rev. B **73**, 214427 (2006).
[22] A. Lavarélo, G. Roux, and N. Laflorencie, Phys. Rev. B **84**, 144407 (2011).
[23] In Ref. 21 (22), the coordination number between sites on opposite legs of the ladder that are connected by antiferromagnetic exchange couplings is three (one).
[24] N. Chepiga, I. Affleck, and F. Mila, Phys. Rev. B **93**, 241108 (2016).
[25] S. R. White, Phys. Rev. Lett. **69**, 2863 (1992).
[26] S. R. White, Phys. Rev. B **48**, 10345 (1993).
[27] Let $l$ be an integer labeling sites along one leg of a two-leg ladder or the sites along a zig-zag path that alternates between the lower and upper legs of the two-leg

ladder. Define the three-site, two-spin operator $\hat{D}_l := \hat{\boldsymbol{S}}_l \cdot \left(\hat{\boldsymbol{S}}_{l+1} - \hat{\boldsymbol{S}}_{l-1}\right)$. The local order parameter for dimer order, $D_l$, is the ground state expectation of $\hat{D}_l$.
[28] J. L. Cardy, Nucl. Phys. B **275**, 200 (1986).
[29] P. Calabrese and J. Cardy, J. Stat. Mech. (2004) P06002.
[30] P. Calabrese and J. Cardy, J. Phys. A **42**, 504005 (2009).
[31] N. Laflorencie, E. S. Sørensen, M.-S. Chang, and I. Affleck, Phys. Rev. Lett. **96**, 100603 (2006).
[32] I. Affleck, N. Laflorencie, and E. S. Srensen, J. Phys. A **42**, 504009 (2009).
[33] J. Cardy and P. Calabrese, J. Stat. Mech. (2010) P04023.
[34] I. Affleck and F. D. M. Haldane, Phys. Rev. B **36**, 5291 (1987).
[35] A. A. Nersesyan, A. O. Gogolin, and F. H. L. Eßler, Phys. Rev. Lett. **81**, 910 (1998).
[36] O. A. Starykh and L. Balents, Phys. Rev. Lett. **93**, 127202 (2004).
[37] O. A. Starykh and L. Balents, Phys. Rev. Lett. **98**, 077205 (2007).
[38] Work in progress.

# Supplementary Material

Po-Hao Huang,[1] Jyong-Hao Chen,[2] Adrian E. Feiguin,[3] Claudio Chamon,[1] and Christopher Mudry[2]
[1]*Department of Physics, Boston University, Boston, MA, 02215, USA*
[2]*Condensed Matter Theory Group, Paul Scherrer Institute, CH-5232 Villigen PSI, Switzerland*
[3]*Department of Physics, Northeastern University, Boston, Massachusetts 02115, USA*

## I. INTRODUCTION

In this Supplementary Material, we are first going to review how the two-leg ladder depicted in Fig. 1 can be described at low energies as a perturbed conformal field theory (CFT). We then weakly couple the two-leg ladders as shown in Fig. 2 and derive the corresponding perturbed CFT.

## II. FERMIONIC REPRESENTATION OF A QUANTUM CRITICAL POINT WITH $\widehat{u}(1) \oplus \widehat{su}(2)_1$ AFFINE LIE ALGEBRA SYMMETRY

Let $\psi^{*}_{M,\sigma}$ and $\psi_{M,\sigma}$, where the label $M = L, R$ stands for left- and right-movers while the label $\sigma = 1, 2$ stands for some internal indices, denote a pair of independent Grassmann-valued fields that depend on imaginary-time $\tau \in \mathbb{R}$ and position $x \in \mathbb{R}$. Define the Lagrangian density

$$\mathcal{L}_0 := \Big[\psi^{*}_{L,\sigma}\,(\partial_\tau + \mathrm{i}v\,\partial_x)\,\psi_{L,\sigma} + \psi^{*}_{R,\sigma}\,(\partial_\tau - \mathrm{i}v\,\partial_x)\,\psi_{R,\sigma}\Big]. \tag{2.1a}$$

Here, the summation over the repeated indices $\sigma = 1, 2$ is implied. Define the action

$$S_0 := \int \mathrm{d}\tau \int \mathrm{d}x\, \mathcal{L}_0. \tag{2.1b}$$

Define the partition function

$$Z_0 := \int \mathcal{D}[\psi^{*}, \psi]\; e^{-S_0}. \tag{2.1c}$$

The expectation value of any product of the Grassmann fields is an algebraic function of the arguments of the Grassmann fields that factorizes into left- and right-moving sectors. This fact is a consequence of applying Wick's theorem with the fundamental two-point correlation functions

$$\langle \psi_{M,\sigma}(x_M)\, \psi^{*}_{M',\sigma'}(0)\rangle = \delta_{MM'}\,\delta_{\sigma\sigma'}\,\frac{1}{2\pi}\,\frac{1}{x_M}, \tag{2.2a}$$

where

$$x_M := \begin{cases} v\,\tau + \mathrm{i}x, & \text{if } M = L,\\ v\,\tau - \mathrm{i}x, & \text{if } M = R. \end{cases} \tag{2.2b}$$

The partition function $Z_0$ is invariant under any local linear transformation

$$(U^{(L)}, U^{(R)}) \in U(2)\times U(2) \tag{2.3a}$$

defined by the fundamental rules

$$\begin{aligned} \psi^{*\mathsf{T}}_{M}(x_M) &\mapsto \psi^{*\mathsf{T}}_{M}(x_M)\, U^{(L)\dagger}(x_M),\\ \psi_{M}(x_M) &\mapsto U^{(L)}(x_M)\,\psi_{M}(x_M), \end{aligned} \tag{2.3b}$$

on the Grassmann integration variables ($M = L, R$). The generators of these symmetries are comprised of the left- and right-moving resolved $u(1)$ currents

$$j_M := \psi^{*}_{M,\sigma}\,\delta_{\sigma\sigma'}\,\psi_{M,\sigma'} \tag{2.4a}$$

and of the left- and right-moving resolved $su(2)$ currents

$$J^{a}_{M} := \frac{1}{2}\,\psi^{*}_{M,\sigma}\,\sigma^{a}_{\sigma,\sigma'}\,\psi_{M,\sigma'}. \tag{2.4b}$$

Here, $\sigma^a$ with $a = 1, 2, 3$ are the Pauli matrices and $M = L, R$. We shall also complement the vector $\boldsymbol{\sigma}$ made out of the three Pauli matrices by the unit $2\times 2$ matrix $\sigma^0$.

The operator product expansion (OPE) between any pair of these currents follow from the identities

$$\sigma^a\,\sigma^b = \delta^{ab}\,\sigma^0 + \mathrm{i}\epsilon^{abc}\,\sigma^c, \qquad \mathrm{tr}\,\left(\sigma^a\,\sigma^b\right) = 2\delta^{ab}, \tag{2.5}$$

where $a, b = 1, 2, 3$ and summation over repeated indices is implied. In the right-moving sector, the only non-vanishing ones are

$$j_R(x_R)\,j_R(0) \sim \frac{1}{(2\pi)^2}\,\frac{2}{x_R^2} \tag{2.6a}$$

and

$$J^a_R(x_R)\,J^b_R(0) \sim \frac{1}{(2\pi)^2}\,\frac{\delta^{ab}/2}{x_R^2} + \frac{1}{2\pi}\,\frac{\mathrm{i}\epsilon^{abc}J^c_R(0)}{x_R}. \tag{2.6b}$$

The OPE between left-moving currents are obtained by the substitution of $x_R \equiv v\,\tau - \mathrm{i}x$ with $x_L \equiv v\,\tau + \mathrm{i}x$ in those for the right-moving currents. The OPE between left-and right-moving currents vanish. These OPE define the direct sum over an affine $\widehat{u}(1)$ Lie algebra and an affine $\widehat{su}(2)_1$ Lie algebra.

The symmetries (2.3) imply the Sugawara identities[1]

$$\begin{aligned} T_M &:= \frac{1}{2\pi}\sum_{\sigma=\uparrow,\downarrow}\psi^{*}_{M,\sigma}\,2\partial^{*}_{M}\,\psi_{M,\sigma}\\ &= T_M[\hat{u}(1)] + T_M[\widehat{su}(2)_1], \end{aligned} \tag{2.7a}$$

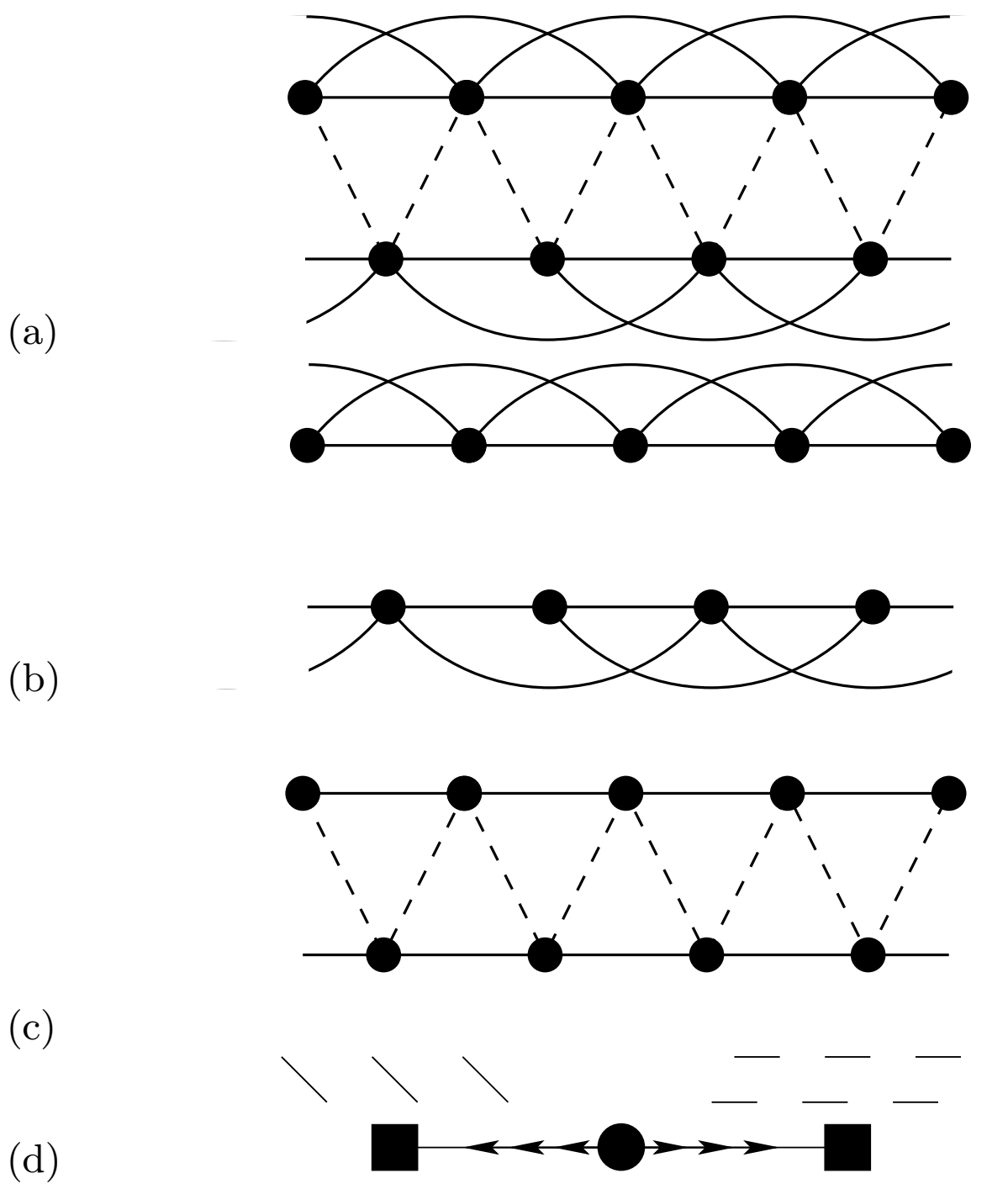

FIG. 1. (a) A two-leg ladder is the set of points represented by the filled black circles on which the quantum spin-1/2 degrees of freedom are localized. The chain of upper (lower) circles defines the upper (lower) leg. The bonds between two consecutive sites on either the upper or lower leg represent the antiferromagnetic exchange coupling $J_1 \geq 0$. The bonds between two next-nearest neighbor sites on either the upper or lower leg represent the antiferromagnetic exchange coupling $J_2 \geq 0$. The dashed bonds across the upper and lower legs represent the antiferromagnetic exchange coupling $J_\vee \geq 0$. (b) When $J_\vee = 0$, the two-leg ladder decouples into two identical $J_1 - J_2$ antiferromagnetic Heisenberg chains. (c) When $J_2 = 0$, the two-leg ladder turns into a single $J_\vee - J_1$ antiferromagnetic Heisenberg chain. (d) One-dimensional phase diagram in parameter space relating the two Majumdar-Ghosh points $J_2 = 0$, $J_1/J_\vee = 1/2$ and $J_\vee = 0$, $J_2/J_1 = 1/2$. The Majumdar-Ghosh points are represented by squares. They realize gapped phases. The gap closes in a continuous fashion at the unstable quantum critical point represented by the filled circle that belongs to the Ising universality class.

where

$$T_M[\hat{u}(1)] = \frac{1}{4} j_M\, j_M, \quad T_M[\widehat{su}(2)_1] = \frac{1}{3}\, \boldsymbol{J}_M \cdot \boldsymbol{J}_M, \tag{2.7b}$$

for components $M = L, R$ of the energy-momentum tensor. Here,

$$\begin{aligned} \frac{\partial}{\partial x_M} &\equiv \partial_M \\ &= \frac{1}{2} \times \begin{cases} \left(v^{-1}\,\partial_\tau - \mathrm{i}\partial_x\right), & \text{if } M = L, \\ \left(v^{-1}\,\partial_\tau + \mathrm{i}\partial_x\right), & \text{if } M = R. \end{cases} \end{aligned} \tag{2.7c}$$

The OPE obeyed by the energy-momentum tensor real-ize the direct sum of two Virasoro algebras, each of which carries the central charge one. Hence, the partition function (2.1) defines a quantum-critical theory with central charge two, where the $\widehat{u}(1)$ sector contributes a central charge of one while the $\widehat{su}(2)_1$ sector contributes another central charge of one.

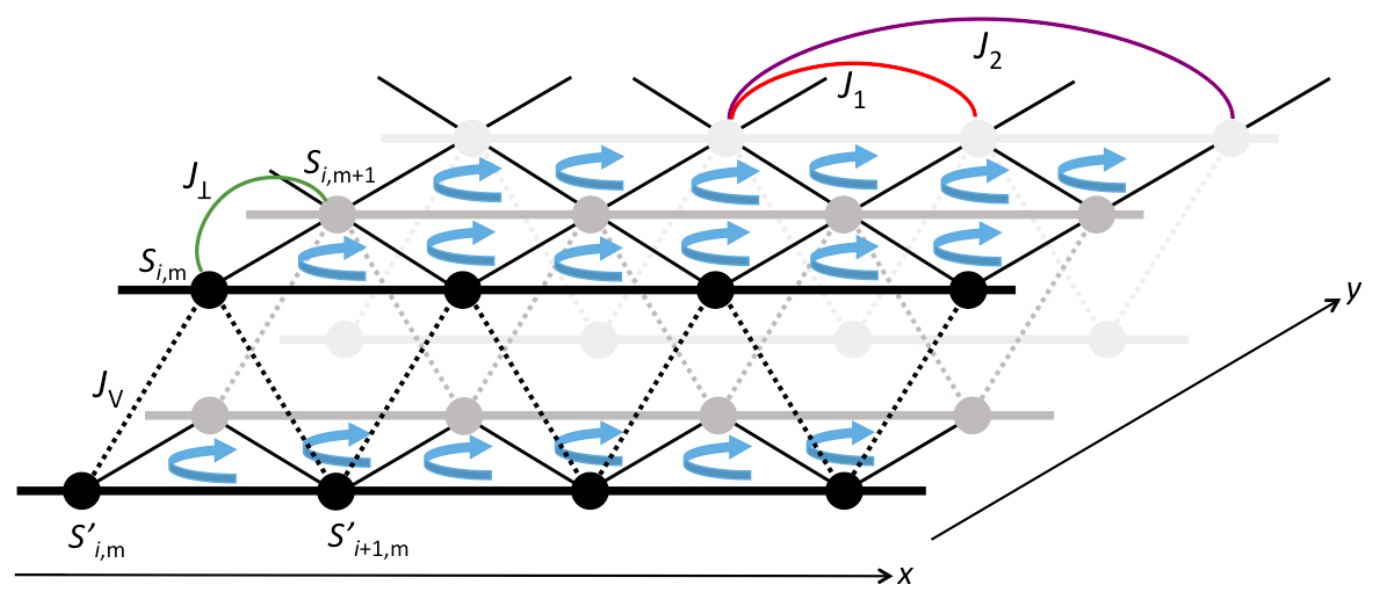

FIG. 2. (Color online) Strongly coupled spin-1/2 two-leg ladders that realize the Ising topological order in two-dimensional space. The intra-ladder couplings $J_1$, $J_2$, and $J_\vee$ are defined in Eq. (3.1). The inter-ladder couplings $J_\perp$ (represented by the green bond) and $\chi_\perp$ (represented by the blue arrows) are defined in Eq. (4.1).

Define the quadratic perturbations

$$\varepsilon_{\mathrm{bs}} := \frac{\mathrm{i}}{2}\left[\psi^*_{R,\sigma}\,\delta_{\sigma\sigma'}\,\psi_{L\sigma'} - (L \leftrightarrow R)\right], \tag{2.8a}$$

and

$$N^a_{\mathrm{bs}} := \frac{1}{2}\left[\psi^*_{R,\sigma}\,\sigma^a_{\sigma\sigma'}\,\psi_{L\sigma'} + (L \leftrightarrow R)\right], \tag{2.8b}$$

where the subscript refers to backscattering. The scaling dimensions of these perturbations are one. However, these perturbations are related to fields carrying anomalous scaling dimensions one half, as we shall demonstrate shortly. Before doing so, however, we observe that the following OPE hold[2–4],

$$\varepsilon_{\rm bs}(\tau,x)\,\varepsilon_{\rm bs}(0,0)\sim\frac{1}{(2\pi)^2\,x_{\rm L}\,x_{\rm R}},\tag{2.9a}$$

$$N^a_{\rm bs}(\tau,x)\,N^b_{\rm bs}(0,0)\sim\frac{\delta^{ab}}{(2\pi)^2\,x_{\rm L}\,x_{\rm R}}\tag{2.9b}$$

$$+\frac{\mathrm{i}\epsilon^{abc}}{2\pi}\left[\frac{J^c_L(0)}{x_R}+\frac{J^c_R(0)}{x_L}\right],\tag{2.9c}$$

$$N^a_{\rm bs}(\tau,x)\,\varepsilon_{\rm bs}(0,0)\sim\frac{\mathrm{i}}{2\pi}\left[\frac{J^a_L(0)}{x_R}-\frac{J^a_R(0)}{x_L}\right],\tag{2.9d}$$

$$J^a_L(x_L)\,N^b_{\rm bs}(0,0)\sim\frac{\mathrm{i}\epsilon^{abc}\,N^c_{\rm bs}(0,0)+\mathrm{i}\delta^{ab}\,\varepsilon_{\rm bs}(0,0)}{4\pi\,x_L},\tag{2.9e}$$

$$J^a_R(x_R)\,N^b_{\rm bs}(0,0)\sim\frac{\mathrm{i}\epsilon^{abc}\,N^c_{\rm bs}(0,0)-\mathrm{i}\delta^{ab}\,\varepsilon_{\rm bs}(0,0)}{4\pi\,x_R},\tag{2.9f}$$

$$J^a_L(x_L)\,\varepsilon_{\rm bs}(0,0)\sim-\frac{\mathrm{i}N^a_{\rm bs}(0,0)}{4\pi\,x_L},\tag{2.9g}$$

$$J^a_R(x_R)\,\varepsilon_{\rm bs}(0,0)\sim+\frac{\mathrm{i}N^a_{\rm bs}(0,0)}{4\pi\,x_R}.\tag{2.9h}$$

We now return to the question of extracting quantum fields with anomalous scaling dimensions one half at this quantum critical point. To this end, we use the representation of the quantum critical theory (2.1) in terms of fields taking values in

$$\begin{aligned}U(2)=&\,U(1)\times SU(2)\\ \equiv&\left\{u\;\middle|\;u=e^{\mathrm{i}\sqrt{2\pi}\phi_c}\,g,\ \phi_c\in[0,\sqrt{2\pi}[,\ g\in SU(2)\right\}.\end{aligned}\tag{2.10}$$

The quantum critical point (2.1) can be represented by the product of two bosonic partition functions

$$Z_0=Z^{U(1)}_0\times Z^{SU(2)}_{\rm WZW}.\tag{2.11a}$$

Here,

$$Z^{U(1)}_0:=\int\mathcal{D}[\phi_{\rm c}]\,e^{-S^{U(1)}_0}\tag{2.11b}$$

is the bosonic partition function for the real-valued and noninteracting scalar field $\phi_{\rm c}$, while

$$Z^{SU(2)}_{\rm WZW}:=\int\mathcal{D}[g]\,e^{-S^{SU(2)}-\Gamma^{SU(2)}_{\rm WZ}}\tag{2.11c}$$

is the Wess-Zumino-Witten (WZW) bosonic partition function for the $SU(2)$-valued matrix field $g$, i.e., an interacting quantum-field theory at criticality. The noninteracting $U(1)$ action is (we momentarily set $v=1$ to save space)

$$S^{U(1)}_0:=\int\mathrm{d}\tau\int\mathrm{d}x\,\mathcal{L}^{U(1)}_0\tag{2.11d}$$

with the Lagrangian density

$$\mathcal{L}^{U(1)}_0:=\frac{1}{2}\left[(\partial_\tau\phi_{\rm c})^2+(\partial_x\phi_{\rm c})^2\right].\tag{2.11e}$$

The interacting WZW action is the sum of (we momentarily set $v=1$ to save space)

$$S^{SU(2)}:=\int\mathrm{d}\tau\int\mathrm{d}x\,\mathcal{L}^{SU(2)}\tag{2.11f}$$

with the local Lagrangian density

$$\mathcal{L}^{SU(2)}=-\frac{1}{4\pi}\mathrm{tr}\left[g^{-1}\left(\partial_\tau g\right)\,g^{-1}\left(\partial_\tau g\right)+(\tau\to x)\right]\tag{2.11g}$$

and the Wess-Zumino action

$$\Gamma^{SU(2)}_{\rm WZ}=\int\mathrm{d}\tau\int\mathrm{d}x\int\limits_0^1\mathrm{d}\xi\times\frac{-\mathrm{i}\,\epsilon^{\mu\nu\rho}}{6\pi}\,\mathrm{tr}\left[\bar g^{-1}\left(\partial_\mu\bar g\right)\,\bar g^{-1}\left(\partial_\nu\bar g\right)\,\bar g^{-1}\left(\partial_\rho\bar g\right)\right],\tag{2.11h}$$

where $\bar g(\tau,x,\xi)$ interpolates between the unit matrix at $\xi=0$ and the field $g(\tau,x)$ at $\xi=1$ and $\partial_\mu\equiv(\partial_\tau,\partial_x,\partial_\xi)$. (We must substitute $\tau$ by $v\,\tau$ to reinstate $v\neq1$.)

The two-point correlation function for the Gaussian field $\phi_{\rm c}$ is

$$\langle\phi_{\rm c}(\tau,x)\;\phi_{\rm c}(0,0)\rangle=-\frac{1}{4\pi}\ln\left(\frac{x_L\,x_R}{\mathfrak{a}^2}\right)\tag{2.12}$$

with $\mathfrak{a}$ a short-distance cutoff. There follows the OPE

$$e^{\mathrm{i}a\,\phi_{\rm c}(\tau,x)}\,e^{\mathrm{i}b\,\phi_{\rm c}(0,0)}\sim\left(\frac{\mathfrak{a}^2}{x_L\,x_R}\right)^{(\Delta_a+\Delta_b-\Delta_{a+b})/2}\times e^{\mathrm{i}(a+b)\,\phi_{\rm c}(0,0)}+\cdots\tag{2.13a}$$

with

$$\Delta_a:=\frac{a^2}{4\pi}\tag{2.13b}$$

the anomalous scaling dimension of the vertex operator $\exp\left(\mathrm{i}a\,\phi_{\rm c}(\tau,x)\right)$. Hence, the vertex operator $\exp\left(\mathrm{i}\sqrt{2\pi}\,\phi_{\rm c}(\tau,x)\right)$ has the anomalous scaling dimension 1/2. The same is true of the matrix-valued field $g\in SU(2)$, since it obeys the OPE

$$g_{\alpha\beta}(\tau,x)\,g_{\alpha'\beta'}(0,0)\sim\mathfrak{a}\,\frac{\epsilon_{\alpha\alpha'}\,\epsilon_{\beta\beta'}}{(x_L\,x_R)^{1/2}}+\cdots\tag{2.14}$$

for $\alpha,\beta,\alpha',\beta'=1,2$ and with $\epsilon_{\alpha\beta}$ the $SU(2)$ Levi-Civita invariant tensor.

To make contact with the fermionic representation (2.1) of the quantum critical point one uses the bosoniza-

tion rules

$$j_L \sim +\frac{\mathrm{i}}{\sqrt{\pi}}\,\partial_L\phi_{\mathrm{c}}, \qquad j_R \sim -\frac{\mathrm{i}}{\sqrt{\pi}}\,\partial_R\phi_{\mathrm{c}}, \tag{2.15a}$$

$$\boldsymbol{J}_L\cdot\frac{\boldsymbol{\sigma}}{2} \sim +\frac{1}{4\pi}(\partial_L g)\,g^{-1}, \quad \boldsymbol{J}_R\cdot\frac{\boldsymbol{\sigma}}{2} \sim -\frac{1}{4\pi}g^{-1}\,(\partial_R g), \tag{2.15b}$$

$$\psi^{*}_{L,\sigma}\,\psi_{R,\sigma'} \sim \frac{1}{2\pi\mathfrak{a}}\,e^{+\mathrm{i}\sqrt{2\pi}\,\phi_{\mathrm{c}}}\,g_{\sigma'\sigma}, \qquad \sigma,\sigma'=1,2. \tag{2.15c}$$

Consequently,

$$\varepsilon_{\mathrm{bs}} \sim \sin(\sqrt{2\pi}\phi_c)\,\varepsilon, \qquad \varepsilon \coloneqq \frac{1}{2\pi\mathfrak{a}}\,\mathrm{tr}\,g, \tag{2.16a}$$

and

$$\boldsymbol{N}_{\mathrm{bs}} \sim \sin(\sqrt{2\pi}\phi_c)\,\boldsymbol{n}, \qquad \boldsymbol{n} \coloneqq \frac{\mathrm{i}}{2\pi\mathfrak{a}}\,\mathrm{tr}\,(g\,\boldsymbol{\sigma}), \tag{2.16b}$$

respectively.

The OPE in two-dimensional Euclidean spacetime between the fields

$$J^a_L \coloneqq +\frac{1}{4\pi}\,\mathrm{tr}\,\left[\sigma^a\,(\partial_L g)\,g^{-1}\right], \qquad a=1,2,3, \tag{2.17a}$$

$$J^a_R \coloneqq -\frac{1}{4\pi}\,\mathrm{tr}\,\left[\sigma^a\,g^{-1}\,(\partial_R g)\right], \qquad a=1,2,3, \tag{2.17b}$$

$$n^a \coloneqq \frac{\mathrm{i}}{2\pi\mathfrak{a}}\,\mathrm{tr}\,(\sigma^a\,g)\,, \qquad a=1,2,3, \tag{2.17c}$$

$$\varepsilon \coloneqq \frac{1}{2\pi\mathfrak{a}}\,\mathrm{tr}\,(\,g), \tag{2.17d}$$

at the $\widehat{su}(2)_1$ quantum critical point follow from the fermionic OPE (2.9) given the identifications (2.15) and (2.16). They are the closed affine $\widehat{su}(2)_1$ algebra

$$J^a_M(x_M)\,J^b_M(0) \sim \frac{1}{(2\pi)^2}\,\frac{\delta^{ab}/2}{x^2_M}+\frac{1}{2\pi}\,\frac{\mathrm{i}\epsilon^{abc}J^c_R(0)}{x_M}, \tag{2.18a}$$

the closed algebras

$$n^a(\tau,x)\,n^b(0,0) \sim \frac{1}{2\pi^2\,\mathfrak{a}}\,\frac{\delta^{ab}}{(x_L\,x_R)^{1/2}}+\cdots \tag{2.18b}$$

and

$$\varepsilon(\tau,x)\,\varepsilon(0,0) \sim \frac{1}{2\pi^2\,\mathfrak{a}}\,\frac{\delta^{ab}}{(x_L\,x_R)^{1/2}}+\cdots\,, \tag{2.18c}$$

with the nonvanishing cross terms

$$J^a_L(x_L)\,n^b(0,0) \sim \frac{\mathrm{i}\epsilon^{abc}\,n^c(0,0)+\mathrm{i}\delta^{ab}\,\varepsilon(0,0)}{4\pi\,x_L}, \tag{2.18d}$$

$$J^a_R(x_R)\,n^b(0,0) \sim \frac{\mathrm{i}\epsilon^{abc}\,n^c(0,0)-\mathrm{i}\delta^{ab}\,\varepsilon(0,0)}{4\pi\,x_R}, \tag{2.18e}$$

$$J^a_L(x_L)\,\varepsilon(0,0) \sim -\frac{\mathrm{i}n^a(0,0)}{4\pi\,x_L}, \tag{2.18f}$$

$$J^a_R(x_R)\,\varepsilon(0,0) \sim +\frac{\mathrm{i}n^a(0,0)}{4\pi\,x_R}. \tag{2.18g}$$

## III. CONTINUUM LIMIT FOR A SINGLE LADDER

The spin-1/2 two-leg ladder depicted in Fig. 1 is a set of two one-dimensional chains, each of which has the lattice spacing $\mathfrak{a}$. Sites along the upper leg of the ladder are labeled by the integer $i$. Sites along the lower leg of the ladder are labeled by the integer $i'$. Each site of the ladder is assigned the generators of the spin-1/2 algebra in the fundamental representation of the group $SU(2)$. The corresponding Hamiltonian is

$$\widehat{H}_{\mathrm{ladder}} \coloneqq \widehat{H}_{\mathrm{leg}}+\widehat{H}'_{\mathrm{leg}}+\widehat{H}_{\mathrm{zig-zag}}. \tag{3.1a}$$

Here,

$$\widehat{H}_{\mathrm{leg}} \coloneqq \sum_{i=1}^{N}\Big(J_1\,\widehat{\boldsymbol{S}}_i\cdot\widehat{\boldsymbol{S}}_{i+1}+J_2\,\widehat{\boldsymbol{S}}_i\cdot\widehat{\boldsymbol{S}}_{i+2}\Big) \tag{3.1b}$$

with $J_1, J_2 \geq 0$ and $\widehat{H}'_{\mathrm{leg}}$ obtained from $\widehat{H}_{\mathrm{leg}}$ with the substitution $\widehat{\boldsymbol{S}}_i \to \widehat{\boldsymbol{S}}'_{i'}$, and

$$\widehat{H}_{\mathrm{zig-zag}} \coloneqq J_{\vee}\sum_{i,i'=1}^{N}\Big(\delta_{i'i}+\delta_{i'(i+1)}\Big)\,\widehat{\boldsymbol{S}}_i\cdot\widehat{\boldsymbol{S}}'_{i'} \tag{3.1c}$$

with $J_{\vee} \geq 0$. Any pair of spin-1/2 operator commute when labeled by two distinct sites of the ladder. Otherwise, they obey the $su(2)$ Lie algebra

$$[\widehat{S}^a_i,\widehat{S}^b_i]=\mathrm{i}\epsilon^{abc}\,\widehat{S}^c_i \tag{3.1d}$$

in the representation

$$\widehat{\boldsymbol{S}}^2_i=\frac{3}{4} \tag{3.1e}$$

for the site $i$, say.

Hamiltonian (3.1a) is invariant under any global $SU(2)$ rotation of all spins, the interchange of the upper and lower legs, and the mirror symmetries centered about a site of one leg and the middle of the bond of the other leg [see Fig. 1(a)].

### A. The case $J_{\vee}=0$

When

$$J_{\vee}=0, \tag{3.2}$$

the ladder Hamiltonian (3.1) decouples into two independent quantum spin-1/2 chains with nearest- and next-nearest-neighbor antiferromagnetic Heisenberg exchange couplings $J_1>0$ and $J_2\geq 0$, respectively [see Fig. 1(b)]. Without loss of generality, we shall consider the Hamiltonian (3.1a) for the upper leg only. The results below apply to the lower leg by adding a prime to all operators and quantum fields.

The phase diagram along the line parameterized by the dimensionless coupling $J_2/J_1 > 0$ consists of the quantum critical segment $0 < J_2/J_1 \leq (J_2/J_1)_{\mathrm{c}}$, the quantum critical end point $(J_2/J_1)_{\mathrm{c}}$, and the gapped phase along the semi-infinite segment $(J_2/J_1)_{\mathrm{c}} < J_2/J_1 < \infty$[5].

Each critical point along the segment $0 < J_2/J_1 \leq (J_2/J_1)_{\mathrm{c}}$ realizes the Wess-Zumino-Witten (WZW) theory defined by the partition function (2.11c) in two-dimensional Euclidean spacetime.

The gapped phase breaks spontaneously the translation symmetry by one lattice spacing of the spin-1/2 chain through the onset of long-ranged dimer order when $J_2/J_1$ becomes larger than $(J_2/J_1)_{\mathrm{c}}$. At the Majumdar-Ghosh point[6], $(J_2/J_1)_{\mathrm{MG}} \equiv 1/2 > (J_2/J_1)_{\mathrm{c}}$, the ground state for an even number of sites $N$ and with periodic boundary conditions (PBC) is two-fold degenerate, i.e., any linear combination of the two valence bond states

$$\prod_{i=1}^{N/2} \frac{1}{\sqrt{2}} \left( |\uparrow,\downarrow\rangle_{2i-1,2i} - |\downarrow,\uparrow\rangle_{2i-1,2i} \right) \tag{3.3a}$$

and

$$\prod_{i=1}^{N/2} \frac{1}{\sqrt{2}} \left( |\uparrow,\downarrow\rangle_{2i,2i+1} - |\downarrow,\uparrow\rangle_{2i,2i+1} \right). \tag{3.3b}$$

This phase diagram can be derived as was done in Ref. 7. First, each spin operator $\widehat{\boldsymbol{S}}_i$ is represented by slave-fermions through

$$\widehat{\boldsymbol{S}}_i = \frac{1}{2}\, \widehat{c}^{\dagger}_i\, \boldsymbol{\sigma}\, \widehat{c}_i, \tag{3.4a}$$

where the only nonvanishing anticommutators at equal times are

$$\{\widehat{c}_{i\alpha}, \widehat{c}^{\dagger}_{i'\alpha'}\} = \delta_{ii'}\,\delta_{\alpha\alpha'} \tag{3.4b}$$

and the local constraint

$$\sum_{\alpha,\beta=1,2} \widehat{c}^{\dagger}_{i\alpha}\,\delta_{\alpha\beta}\,\widehat{c}_{i\beta} = 1 \tag{3.4c}$$

holds. Second, the partition function for the $J_1$-$J_2$ quantum spin-1/2 antiferromagnetic Heisenberg chain can be represented exactly by a lattice gauge theory for fermions and bosons obeying a local $SU(2)$ gauge symmetry. Third, at low energies, this lattice gauge theory can be approximated by quantum chromodynamics in two-dimensional Minkowski spacetime (QCD$_2$) perturbed by some four fermion interactions. More precisely, QCD$_2$ is defined with the $SU(2)$ gauge group in the limit for which the gauge coupling is infinitely strong and with the matter fields in the fundamental representation of $SU(2)$, i.e., the quantum critical theory (2.1) subject to the local constraints

$$\boldsymbol{J}_L(x_L) + \boldsymbol{J}_R(x_R) = 0. \tag{3.5}$$

In turn, this low-energy effective theory can be bosonized using functional non-Abelian bosonization[7]. The outcome is the partition function with the $\widehat{su}(2)_1$ WZW *action perturbed by the addition of the term*[7]

$$\left(g^{(1)}_{\mathrm{bs}} - g^{(2)}_{\mathrm{bs}}\right) \int \mathrm{d}t \int \mathrm{d}x\, \boldsymbol{J}_L(v\,t+x)\cdot \boldsymbol{J}_R(v\,t-x) \tag{3.6a}$$

*in two-dimensional Minkowski spacetime*. Observe that this term translates into the current-current interaction

$$\begin{aligned}\widehat{V}_{\mathrm{bs}} :=& -\left(g^{(1)}_{\mathrm{bs}} - g^{(2)}_{\mathrm{bs}}\right) \int \mathrm{d}x\, \widehat{\boldsymbol{J}}_L(v\,t+x)\cdot \widehat{\boldsymbol{J}}_R(v\,t-x)\\ \equiv& \, g_{\mathrm{bs}} \int \mathrm{d}x\, \widehat{\boldsymbol{J}}_L(v\,t+x)\cdot \widehat{\boldsymbol{J}}_R(v\,t-x)\end{aligned} \tag{3.6b}$$

in the Hamiltonian picture of quantum-field theory. Here,

$$0 < g^{(1)}_{\mathrm{bs}} \propto J_1, \qquad 0 < g^{(2)}_{\mathrm{bs}} \propto J_2. \tag{3.6c}$$

The quantum critical regime corresponds to this perturbation being marginally irrelevant,

$$g^{(1)}_{\mathrm{bs}} > g^{(2)}_{\mathrm{bs}}. \tag{3.7a}$$

The gapped regime corresponds to this perturbation being marginally relevant,

$$g^{(1)}_{\mathrm{bs}} < g^{(2)}_{\mathrm{bs}}. \tag{3.7b}$$

The spin operators that were defined on the sites $i$ from the one-dimensional chain with the lattice spacing $\mathfrak{a}$ are encoded in the effective low-energy quantum-field theory by the following quantum fields. If

$$i\,(2\mathfrak{a}) \to x, \qquad N\,\mathfrak{a} \to L, \tag{3.8a}$$

then

$$\widehat{\boldsymbol{S}}_{2i} \to (2\mathfrak{a})\,\left[\widehat{\boldsymbol{m}}(x) + \widehat{\boldsymbol{n}}(x)\right], \tag{3.8b}$$

$$\widehat{\boldsymbol{S}}_{2i+1} \to (2\mathfrak{a})\,\left[\widehat{\boldsymbol{m}}(x) - \widehat{\boldsymbol{n}}(x)\right], \tag{3.8c}$$

for all sites $i = 1, \cdots, N/2$ of the upper leg, assuming that $N$ is even. By combining the normalization (3.1e) with Eqs. (3.8b) and (3.8c), we find the constraint

$$\widehat{\boldsymbol{m}}^2(x) + \widehat{\boldsymbol{n}}^2(x) = \mathrm{C}(\mathfrak{a}) > 0 \tag{3.9}$$

for all $x$, where the value of the positive number $\mathrm{C}(\mathfrak{a})$ is fixed by the representation 1/2 of the $su(2)$ algebra obeyed by the spin operators $\widehat{\boldsymbol{S}}_i$.

By definition

$$\widehat{\boldsymbol{S}}_{2i} + \widehat{\boldsymbol{S}}_{2i+1} \to 2\times(2\mathfrak{a})\,\widehat{\boldsymbol{m}}(x) \tag{3.10a}$$

while

$$\widehat{\boldsymbol{S}}_{2i} - \widehat{\boldsymbol{S}}_{2i+1} \to 2\times(2\mathfrak{a})\,\widehat{\boldsymbol{n}}(x). \tag{3.10b}$$

It then follows that the quantum fields $\widehat{\boldsymbol{m}}(x)$ and $\widehat{\boldsymbol{n}}(x)$ commute at equal time,

$$[\widehat{\boldsymbol{m}}(x), \widehat{\boldsymbol{n}}(y)] = 0 \tag{3.11}$$

for all $x$ and $y$ from $[0, L]$. Furthermore, if we assume that the quantum fields $\widehat{\boldsymbol{m}}(x)$ and $\widehat{\boldsymbol{n}}(x)$ vary smoothly relative to the length scale $2\mathfrak{a}$, we may then interpret the former vector of quantum fields as encoding smooth fluctuations of the uniform magnetization and the latter vector of quantum fields as encoding smooth fluctuations of the staggered magnetization.

Finally, the decomposition

$$\widehat{\boldsymbol{m}}(\tau, x) = \widehat{\boldsymbol{J}}_L(v\,\tau + \mathrm{i}x) + \widehat{\boldsymbol{J}}_R(v\,\tau - \mathrm{i}x), \tag{3.12a}$$

and the identifications

$$\widehat{\boldsymbol{J}}_M(x_M) \to \boldsymbol{J}_M(x_M), \qquad \widehat{\boldsymbol{n}}(\tau, x) \to \boldsymbol{n}(\tau, x) \tag{3.12b}$$

hold between the operators $\widehat{\boldsymbol{J}}_M$ with $M = L, R$ and $\widehat{\boldsymbol{n}}$ in the imaginary-time Heisenberg picture and the bosonic fields (2.17a), (2.17b), and (2.17c), respectively, entering the operator content of the conformal field theory with the affine $\widehat{su}(2)_1$ algebra derived from the fundamental OPE (2.14).

Having established the nature of the line of quantum critical points along the segment $0 < J_2/J_1 \leq (J_2/J_1)_{\mathrm{c}}$ and the dictionary relating the spins to the operator content at criticality, we can construct the continuum limit for the perturbations of the critical segment

$$J_\vee = 0, \qquad 0 < J_2/J_1 \leq (J_2/J_1)_{\mathrm{c}} \tag{3.13}$$

in the parameter space for the two-leg ladder.

### B. The case $J_\vee \neq 0$

To obtain the naive continuum limit of the Hamiltonian (3.1c), treated as a perturbation to the critical segment $0 < J_2/J_1 \leq (J_2/J_1)_{\mathrm{c}}$, we first write

$$\begin{aligned}\widehat{H}_{J_\vee} := J_\vee \sum_{i=1}^{N/2} \widehat{\boldsymbol{S}}_{2i} \cdot \left(\widehat{\boldsymbol{S}}'_{2i} + \widehat{\boldsymbol{S}}'_{2i+1}\right) \\ + J_\vee \sum_{i=1}^{N/2} \widehat{\boldsymbol{S}}_{2i+1} \cdot \left(\widehat{\boldsymbol{S}}'_{2i+1} + \widehat{\boldsymbol{S}}'_{2i+2}\right),\end{aligned} \tag{3.14}$$

where we have assumed that $N$ is an even integer and imposed PBC. If we insert the decomposition (3.8) into the chain-like Hamiltonian (3.14), we get

$$\begin{aligned}\widehat{H}_{J_\vee} \to (2\mathfrak{a})\,J_\vee \int_0^L \mathrm{d}x\, [\widehat{\boldsymbol{m}}(x) + \widehat{\boldsymbol{n}}(x)] \cdot \Big[\widehat{\boldsymbol{m}}'(x) + \underline{\widehat{\boldsymbol{n}}'(x)} + \widehat{\boldsymbol{m}}'(x) - \underline{\widehat{\boldsymbol{n}}'(x)}\Big] \\ + (2\mathfrak{a})\,J_\vee \int_0^L \mathrm{d}x\, [\widehat{\boldsymbol{m}}(x) - \widehat{\boldsymbol{n}}(x)] \cdot \Big[\widehat{\boldsymbol{m}}'(x) - \underline{\widehat{\boldsymbol{n}}'(x)} + \widehat{\boldsymbol{m}}'(x + 2\mathfrak{a}) + \underline{\widehat{\boldsymbol{n}}'(x + 2\mathfrak{a})}\Big].\end{aligned} \tag{3.15}$$

To leading order in an expansion in powers of $(2\mathfrak{a})$,

$$\begin{aligned}\widehat{H}_{J_\vee} \to (2\mathfrak{a})\,J_\vee \int_0^L \mathrm{d}x\, [\widehat{\boldsymbol{m}}(x) + \widehat{\boldsymbol{n}}(x)] \cdot [2 \times \widehat{\boldsymbol{m}}'(x)] \\ + (2\mathfrak{a})\,J_\vee \int_0^L \mathrm{d}x\, [\widehat{\boldsymbol{m}}(x) - \widehat{\boldsymbol{n}}(x)] \cdot \left[2 \times \widehat{\boldsymbol{m}}'(x) + (2\mathfrak{a})\,\partial_x \widehat{\boldsymbol{m}}'(x) + (2\mathfrak{a})\,\partial_x \widehat{\boldsymbol{n}}'(x) + \mathcal{O}\big((2\mathfrak{a})^2\big)\right].\end{aligned} \tag{3.16}$$

Hence, the segment of quantum criticality (3.13) in parameter space is perturbed by

$$\widehat{H}^{\vee}_{\mathrm{per}} := \int_0^L \mathrm{d}x\, \left(\widehat{W}^{\vee}_{jj} + \widehat{W}^{\vee}_{\mathrm{tw}} + \cdots\right)(x), \tag{3.17a}$$

where

$$\widehat{W}^{\vee}_{jj}(x) = +\,g_\vee \left[\widehat{\boldsymbol{J}}_L(x) + \widehat{\boldsymbol{J}}_R(x)\right] \cdot \left[\widehat{\boldsymbol{J}}'_L(x) + \widehat{\boldsymbol{J}}'_R(x)\right] \tag{3.17b}$$

is a marginally relevant perturbation[8] with the coupling

$$g_\vee = 4 \times v_\vee > 0, \qquad v_\vee := (2\mathfrak{a})\,J_\vee, \tag{3.17c}$$

and

$$\widehat{W}^{\vee}_{\mathrm{tw}} = -\,g_{\mathrm{tw}}\, \widehat{\boldsymbol{n}}(x) \cdot \partial_x \widehat{\boldsymbol{n}}'(x) \tag{3.17d}$$

is a marginal interaction[9] with the coupling

$$g_{\mathrm{tw}} = (2\mathfrak{a})^2\, v_\vee > 0, \tag{3.17e}$$

and $\cdots$ represents irrelevant local perturbations.

Observe that

$$\begin{aligned}\widehat{V}_{\rm bs}(x)+\widehat{V}'_{\rm bs}(x)+\widehat{W}^{\vee}_{jj}(x)=&\,g_{\rm bs}\left[\widehat{\boldsymbol{J}}_L(x)\cdot\widehat{\boldsymbol{J}}_R(x)+\widehat{\boldsymbol{J}}'_L(x)\cdot\widehat{\boldsymbol{J}}'_R(x)\right]+g_{\vee}\left[\widehat{\boldsymbol{J}}_L(x)+\widehat{\boldsymbol{J}}_R(x)\right]\cdot\left[\widehat{\boldsymbol{J}}'_L(x)+\widehat{\boldsymbol{J}}'_R(x)\right]\\ =&\,g_{\rm bs}\left[\widehat{\boldsymbol{J}}_L(x)\cdot\widehat{\boldsymbol{J}}_R(x)+\widehat{\boldsymbol{J}}'_L(x)\cdot\widehat{\boldsymbol{J}}'_R(x)\right]+g_{\vee}\left[\widehat{\boldsymbol{J}}_L(x)\cdot\widehat{\boldsymbol{J}}'_R(x)+\widehat{\boldsymbol{J}}_R(x)\cdot\widehat{\boldsymbol{J}}'_L(x)\right]\\ &+g_{\vee}\left[\widehat{\boldsymbol{J}}_L(x)\cdot\widehat{\boldsymbol{J}}'_L(x)+\widehat{\boldsymbol{J}}_R(x)\cdot\widehat{\boldsymbol{J}}'_R(x)\right].\end{aligned}\tag{3.18}$$

Define

$$\widehat{\boldsymbol{K}}_L(x)\cdot\widehat{\boldsymbol{K}}_R(x)=\left[\widehat{\boldsymbol{J}}_L(x)+\widehat{\boldsymbol{J}}'_L(x)\right]\cdot\left[\widehat{\boldsymbol{J}}_R(x)+\widehat{\boldsymbol{J}}'_R(x)\right],\tag{3.19a}$$

where

$$\widehat{\boldsymbol{K}}_M(x):=\widehat{\boldsymbol{J}}_M(x)+\widehat{\boldsymbol{J}}'_M(x),\qquad M=L,R.\tag{3.19b}$$

If

$$g_{\rm bs}=g_{\vee}\equiv g_{{\rm bs}=\vee},\tag{3.20}$$

then

$$\widehat{V}_{\rm bs}(x)+\widehat{V}'_{\rm bs}(x)+\widehat{W}^{\vee}_{jj}(x)=g_{{\rm bs}=\vee}\,\widehat{\boldsymbol{K}}_L(x)\cdot\widehat{\boldsymbol{K}}_R(x)+g_{{\rm bs}=\vee}\left[\widehat{\boldsymbol{J}}_L(x)\cdot\widehat{\boldsymbol{J}}'_L(x)+\widehat{\boldsymbol{J}}_R(x)\cdot\widehat{\boldsymbol{J}}'_R(x)\right].\tag{3.21}$$

Ising quantum criticality of the two-leg ladder (3.1) is the consequence of the flow to strong coupling of the local current-current interaction $\left(\widehat{\boldsymbol{K}}_L\cdot\widehat{\boldsymbol{K}}_R\right)(x)$ when it is the only runaway flow from the $\widehat{su}(2)_1\oplus\widehat{su}(2)_1$ quantum critical point. The Ising quantum critical point realizes the coset theory with the affine Lie algebra $\widehat{su}(2)_1\oplus\widehat{su}(2)_1/\widehat{su}(2)_2$, whose central charge is 1/2.

## IV. CONTINUUM LIMIT FOR COUPLED LADDERS

We consider $n$ two-leg ladders labeled by $\mathtt{m}=1,\cdots,n$. They interact through the perturbations $\widehat{H}_{\triangle}$ and $\widehat{H}'_{\triangle}$, where

$$\begin{aligned}\widehat{H}_{\triangle}:=\frac{\chi_{\perp}}{2}\sum_{i=1}^{N}\sum_{\mathtt{m}=1}^{n-1}\Big[&\widehat{\boldsymbol{S}}_{i,\mathtt{m}+1}\cdot\left(\widehat{\boldsymbol{S}}_{i+1,\mathtt{m}}\wedge\widehat{\boldsymbol{S}}_{i,\mathtt{m}}\right)\\ &+\widehat{\boldsymbol{S}}_{i+1,\mathtt{m}}\cdot\left(\widehat{\boldsymbol{S}}_{i,\mathtt{m}+1}\wedge\widehat{\boldsymbol{S}}_{i+1,\mathtt{m}+1}\right)\Big]\end{aligned}\tag{4.1a}$$

with $\chi_{\perp}$ real valued and $\widehat{H}'_{\triangle}$ that follows from $\widehat{H}_{\triangle}$ by the substitution $\widehat{\boldsymbol{S}}_{i,\mathtt{m}}\to\widehat{\boldsymbol{S}}'_{i',\mathtt{m}}$, and the perturbations $\widehat{H}_{\square}$ and $\widehat{H}'_{\square}$, where

$$\begin{aligned}\widehat{H}_{\square}:=\sum_{i=1}^{N}\sum_{\mathtt{m}=1}^{n-1}\Big[&J_{\perp}\,\widehat{\boldsymbol{S}}_{i,\mathtt{m}}\cdot\widehat{\boldsymbol{S}}_{i,\mathtt{m}+1}\\ &+\left(J_{/}\,\widehat{\boldsymbol{S}}_{i,\mathtt{m}}\cdot\widehat{\boldsymbol{S}}_{i+1,\mathtt{m}+1}+J_{\backslash}\,\widehat{\boldsymbol{S}}_{i,\mathtt{m}+1}\cdot\widehat{\boldsymbol{S}}_{i+1,\mathtt{m}}\right)\Big]\end{aligned}\tag{4.1b}$$

with $J_{\perp},J_{/},J_{\backslash}>0$ and $\widehat{H}'_{\square}$ obtained from $\widehat{H}_{\square}$ by the substitution $\widehat{\boldsymbol{S}}_{i,\mathtt{m}}\to\widehat{\boldsymbol{S}}'_{i',\mathtt{m}}$. Figure 2 depicts the special case of

$$J_{\perp}=J_{\backslash},\qquad J_{/}=0,\tag{4.2}$$

which is considered in Eq. (6) from the main text.

The continuum limit of the three-spin interaction $\widehat{H}_{\triangle}$ (4.1a) is derived in Appendix A. The result is the local polynomial

$$\widehat{V}_{\triangle}:=g_{\chi_{\perp}}\sum_{\mathtt{m}=1}^{n-1}\left(\widehat{\boldsymbol{J}}_{L,\mathtt{m}}\cdot\widehat{\boldsymbol{J}}_{R,\mathtt{m}+1}-\widehat{\boldsymbol{J}}_{R,\mathtt{m}}\cdot\widehat{\boldsymbol{J}}_{L,\mathtt{m}+1}\right)+\cdots,\tag{4.3a}$$

where

$$g_{\chi_{\perp}}=2\times(2\mathfrak{a})\,\frac{\chi_{\perp}}{\pi},\tag{4.3b}$$

and $\cdots$ denotes irrelevant local perturbations. Similarly, the continuum limit of $\widehat{H}'_{\triangle}$ delivers the local polynomial $\widehat{V}'_{\triangle}$ that follows from substituting $\widehat{\boldsymbol{J}}_{M,\mathtt{m}}(x)\to\widehat{\boldsymbol{J}}'_{M,\mathtt{m}}(x)$ for $M=L,R$ in $\widehat{V}_{\triangle}$.

The naive continuum limit of $\widehat{H}_{\square}$ defined by Eq. (4.1b) is derived in Appendix B. The result is the local polyno-

mial

$$\widehat{V}_{\square} := \sum_{\mathtt{m}=1}^{n-1} \Bigg[ \sum_{\mathtt{m}'=1}^{n-2} g_{jj}^{(\mathtt{m}')} \left( \widehat{\boldsymbol{J}}_{L,\mathtt{m}} \cdot \widehat{\boldsymbol{J}}_{L,\mathtt{m}+\mathtt{m}'} + \widehat{\boldsymbol{J}}_{R,\mathtt{m}} \cdot \widehat{\boldsymbol{J}}_{R,\mathtt{m}+\mathtt{m}'} \right) + \sum_{\mathtt{m}'=1}^{n-2} g_{jj}^{(\mathtt{m}')} \left( \widehat{\boldsymbol{J}}_{L,\mathtt{m}} \cdot \widehat{\boldsymbol{J}}_{R,\mathtt{m}+\mathtt{m}'} + \widehat{\boldsymbol{J}}_{R,\mathtt{m}} \cdot \widehat{\boldsymbol{J}}_{L,\mathtt{m}+\mathtt{m}'} \right) + \sum_{\mathtt{m}'=1}^{n-2} \left( g_{nn}^{(\mathtt{m}')} \,\widehat{\boldsymbol{n}}_{\mathtt{m}} \cdot \widehat{\boldsymbol{n}}_{\mathtt{m}+\mathtt{m}'} + g_{\varepsilon\varepsilon}^{(\mathtt{m}')} \,\widehat{\varepsilon}_{\mathtt{m}} \,\widehat{\varepsilon}_{\mathtt{m}+\mathtt{m}'} \right) + \sum_{\mathtt{m}'=1}^{n-2} g_{\mathrm{tw}}^{(\mathtt{m}')} \left( \widehat{\boldsymbol{n}}_{\mathtt{m}} \cdot \partial_x \widehat{\boldsymbol{n}}_{\mathtt{m}+\mathtt{m}'} \right) \Bigg] \tag{4.4a}$$

up to irrelevant local perturbations. Here, the bare values of the couplings are

$$g_{jj}^{(\mathtt{m}')} \equiv \left( g_{jj}^{\perp} + g_{jj}^{/} + g_{jj}^{\backslash} \right) \delta_{\mathtt{m}'1} = 2 \times (2\mathfrak{a}) \left( J_{\perp} + J_{/} + J_{\backslash} \right) \delta_{\mathtt{m}'1}, \tag{4.4b}$$

$$g_{nn}^{(\mathtt{m}')} \equiv \left( g_{nn}^{\perp} + g_{nn}^{/} + g_{nn}^{\backslash} \right) \delta_{\mathtt{m}'1} = 2 \times (2\mathfrak{a}) \left( J_{\perp} - J_{/} - J_{\backslash} \right) \delta_{\mathtt{m}'1}, \tag{4.4c}$$

$$g_{\varepsilon\varepsilon}^{(\mathtt{m}')} = 0, \tag{4.4d}$$

$$g_{\mathrm{tw}}^{(\mathtt{m}')} \equiv \left( g_{\mathrm{tw}}^{/} - g_{\mathrm{tw}}^{\backslash} \right) \delta_{\mathtt{m}'1} = (2\mathfrak{a})^2 \left( -J_{/} + J_{\backslash} \right) \delta_{\mathtt{m}'1}. \tag{4.4e}$$

We note that the initial value of the twist coupling vanishes[10] if

$$J_{/} = J_{\backslash}. \tag{4.5}$$

The continuum limit of $\widehat{H}'_{\square}$ can be obtained by the substitutions $\widehat{\boldsymbol{J}}_{M,\mathtt{m}}(x) \to \widehat{\boldsymbol{J}}'_{M,\mathtt{m}}(x)$ with $M = L, R$, $\widehat{\boldsymbol{n}}_{\mathtt{m}}(x) \to \widehat{\boldsymbol{n}}'_{\mathtt{m}}(x)$, and $\widehat{\varepsilon}_{\mathtt{m}}(x) \to \widehat{\varepsilon}'_{\mathtt{m}}(x)$ in $\widehat{V}_{\square}$. The result is the local polynomial $\widehat{V}'_{\square}$.

The condition

$$J_{\perp} = J_{\backslash}, \qquad J_{/} = 0 \tag{4.6}$$

forbids the presence of any relevant perturbation of the form $\widehat{\boldsymbol{n}}_{\mathtt{m}} \cdot \widehat{\boldsymbol{n}}_{\mathtt{m}+\mathtt{m}'}$ and $\widehat{\varepsilon}_{\mathtt{m}} \,\widehat{\varepsilon}_{\mathtt{m}+\mathtt{m}'}$ with $\mathtt{m}'$ an odd integer by the symmetry under reflection in a plane perpendicular to the chains[11,12].

Addition of the contributions to the current-current back-scattering from the local perturbations $\widehat{V}_{\triangle}$, $\widehat{V}_{\square}$, $\widehat{V}'_{\triangle}$, and $\widehat{V}'_{\square}$ [see Eqs. (4.3) and (4.4)] gives the local current-current perturbation

$$\widehat{V}_{\triangle,\square\,\mathrm{bs}} := \sum_{\mathtt{m}=1}^{n-1} \Big[ g_{+} \left( \widehat{\boldsymbol{J}}_{L,\mathtt{m}} \cdot \widehat{\boldsymbol{J}}_{R,\mathtt{m}+1} + \widehat{\boldsymbol{J}}'_{L,\mathtt{m}} \cdot \widehat{\boldsymbol{J}}'_{R,\mathtt{m}+1} \right) + g_{-} \left( \widehat{\boldsymbol{J}}_{R,\mathtt{m}} \cdot \widehat{\boldsymbol{J}}_{L,\mathtt{m}+1} + \widehat{\boldsymbol{J}}'_{R,\mathtt{m}} \cdot \widehat{\boldsymbol{J}}'_{L,\mathtt{m}+1} \right) \Big] \tag{4.7a}$$

with the couplings

$$g_{\pm} := g_{jj}^{(1)} \pm g_{\chi_{\perp}} = 2 \times (2\mathfrak{a}) \left( J_{\perp} + J_{/} + J_{\backslash} \right) \pm 2 \times (2\mathfrak{a}) \frac{\chi_{\perp}}{\pi}. \tag{4.7b}$$

Given the initial values $g_{+} > 0$ and $g_{-} \leq 0$, $g_{+}$ flows to strong coupling while $g_{-}$ flows to zero. Conversely, given the initial values $g_{-} > 0$ and $g_{+} \leq 0$, it is $g_{-}$ that flows to strong coupling while it is $g_{+}$ that flows to zero. The strong coupling fixed points that we seek require that

$$\mathrm{sgn} \left( g_{+} \times g_{-} \right) < 0. \tag{4.8}$$

Furthermore, under the condition (4.6), if we choose

$$\frac{\chi_{\perp}}{\pi} = 2 J_{\perp}, \tag{4.9}$$

then

$$g_{+} \equiv g_{jj} = 8 \times (2\mathfrak{a}) \, J_{\perp}, \qquad g_{-} = 0. \tag{4.10}$$

This is the case considered in Eq. (5) from the main text.

## Appendix A: Continuum limit of the three spin interactions (4.1a)

The perturbation $\widehat{H}_{\triangle}$ defined by Eq. (4.1a) couples two consecutive two-leg ladders in such a way that it breaks time-reversal symmetry. The coupling $\widehat{H}_{\triangle}$ involves two term per square plaquette defined by the vertices $(i,\mathtt{m})$, $(i+1,\mathtt{m})$, $(i,\mathtt{m}+1)$, $(i+1,\mathtt{m}+1)$. We shall take the naive continuum limit arising from each term separately.

We write

$$\begin{aligned}\widehat{H}_{\triangle_1} &:= \frac{\chi_\perp}{2}\sum_{i=1}^{N}\sum_{\mathtt{m}=1}^{n-1}\epsilon^{abc}\widehat{S}^a_{i,\mathtt{m}+1}\,\widehat{S}^b_{i+1,\mathtt{m}}\,\widehat{S}^c_{i,\mathtt{m}} \\ &= \frac{\chi_\perp}{2}\sum_{i=1}^{N/2}\sum_{\mathtt{m}=1}^{n-1}\epsilon^{abc}\left(\widehat{S}^a_{2i,\mathtt{m}+1}\,\widehat{S}^b_{2i+1,\mathtt{m}}\,\widehat{S}^c_{2i,\mathtt{m}}\right) \\ &\quad+\frac{\chi_\perp}{2}\sum_{i=1}^{N/2}\sum_{\mathtt{m}=1}^{n-1}\epsilon^{abc}\left(\widehat{S}^a_{2i+1,\mathtt{m}+1}\,\widehat{S}^b_{2i+2,\mathtt{m}}\,\widehat{S}^c_{2i+1,\mathtt{m}}\right),\end{aligned}\tag{A1}$$

where we have assumed that $N$ is an even integer and imposed PBC. If we insert the decomposition (3.8b) into the Hamiltonian (A1), we get

$$\begin{aligned}\widehat{H}_{\triangle_1} \to &\frac{\chi_\perp}{2}\sum_{\mathtt{m}=1}^{n-1}\int_0^L \mathrm{d}x\,\epsilon^{abc}(2\mathfrak{a})^2\left[\widehat{m}^a_{\mathtt{m}+1}(x)+\widehat{n}^a_{\mathtt{m}+1}(x)\right]\left[\widehat{m}^b_{\mathtt{m}}(x)-\widehat{n}^b_{\mathtt{m}}(x)\right]\left[\widehat{m}^c_{\mathtt{m}}(x)+\widehat{n}^c_{\mathtt{m}}(x)\right] \\ &+\frac{\chi_\perp}{2}\sum_{\mathtt{m}=1}^{n-1}\int_0^L \mathrm{d}x\,\epsilon^{abc}(2\mathfrak{a})^2\left[\widehat{m}^a_{\mathtt{m}+1}(x)-\widehat{n}^a_{\mathtt{m}+1}(x)\right]\underline{\left[\widehat{m}^b_{\mathtt{m}}(x+2\mathfrak{a})+\widehat{n}^b_{\mathtt{m}}(x+2\mathfrak{a})\right]}\left[\widehat{m}^c_{\mathtt{m}}(x)-\widehat{n}^c_{\mathtt{m}}(x)\right].\end{aligned}\tag{A2}$$

To leading order in an expansion in powers of $(2\mathfrak{a})$,

$$\begin{aligned}\widehat{H}_{\triangle_1} \to &\frac{\chi_\perp}{2}\sum_{\mathtt{m}=1}^{n-1}\int_0^L \mathrm{d}x\,\epsilon^{abc}(2\mathfrak{a})^2\left[\widehat{m}^a_{\mathtt{m}+1}(x)+\widehat{n}^a_{\mathtt{m}+1}(x)\right]\left[\widehat{m}^b_{\mathtt{m}}(x)-\widehat{n}^b_{\mathtt{m}}(x)\right]\left[\widehat{m}^c_{\mathtt{m}}(x)+\widehat{n}^c_{\mathtt{m}}(x)\right] \\ &+\frac{\chi_\perp}{2}\sum_{\mathtt{m}=1}^{n-1}\int_0^L \mathrm{d}x\,\epsilon^{abc}(2\mathfrak{a})^2\left[\widehat{m}^a_{\mathtt{m}+1}(x)-\widehat{n}^a_{\mathtt{m}+1}(x)\right]\underline{\left[\widehat{m}^b_{\mathtt{m}}(x)+\widehat{n}^b_{\mathtt{m}}(x)\right]}\left[\widehat{m}^c_{\mathtt{m}}(x)-\widehat{n}^c_{\mathtt{m}}(x)\right] \\ &+\frac{\chi_\perp}{2}\sum_{\mathtt{m}=1}^{n-1}\int_0^L \mathrm{d}x\,\epsilon^{abc}(2\mathfrak{a})^3\left[\widehat{m}^a_{\mathtt{m}+1}(x)-\widehat{n}^a_{\mathtt{m}+1}(x)\right]\underline{\left[\partial_x\widehat{m}^b_{\mathtt{m}}(x)+\partial_x\widehat{n}^b_{\mathtt{m}}(x)\right]}\left[\widehat{m}^c_{\mathtt{m}}(x)-\widehat{n}^c_{\mathtt{m}}(x)\right],\end{aligned}\tag{A3}$$

where the first and the second line can be further simplified,

$$\begin{aligned}\widehat{H}_{\triangle_1} \to &\frac{\chi_\perp}{2}\sum_{\mathtt{m}=1}^{n-1}\int_0^L \mathrm{d}x\,\epsilon^{abc}\,2\times(2\mathfrak{a})^2\Big[\widehat{m}^a_{\mathtt{m}+1}(x)\,\widehat{m}^b_{\mathtt{m}}(x)\,\widehat{m}^c_{\mathtt{m}}(x)-\widehat{m}^a_{\mathtt{m}+1}(x)\,\widehat{n}^b_{\mathtt{m}}(x)\,\widehat{n}^c_{\mathtt{m}}(x) \\ &\qquad+\widehat{n}^a_{\mathtt{m}+1}(x)\,\widehat{m}^b_{\mathtt{m}}(x)\,\widehat{n}^c_{\mathtt{m}}(x)-\widehat{n}^a_{\mathtt{m}+1}(x)\,\widehat{n}^b_{\mathtt{m}}(x)\,\widehat{m}^c_{\mathtt{m}}(x)\Big] \\ &+\frac{\chi_\perp}{2}\sum_{\mathtt{m}=1}^{n-1}\int_0^L \mathrm{d}x\,\epsilon^{abc}(2\mathfrak{a})^3\left[\widehat{m}^a_{\mathtt{m}+1}(x)-\widehat{n}^a_{\mathtt{m}+1}(x)\right]\left[\partial_x\widehat{m}^b_{\mathtt{m}}(x)+\partial_x\widehat{n}^b_{\mathtt{m}}(x)\right]\left[\widehat{m}^c_{\mathtt{m}}(x)-\widehat{n}^c_{\mathtt{m}}(x)\right].\end{aligned}\tag{A4}$$

To proceed, we need to point-splitting pairs of operators sitting at the same position $x$ in the same bundle $\mathtt{m}$. After point-splitting, we shall use the OPE (2.18) so as to reduce the point-split pair of operators to either a $\mathbb{C}$ number or a single operator. We treat such pairs of point-split operators one by one. First,

$$\begin{aligned}\epsilon^{abc}\,\widehat{m}^b_{\mathtt{m}}(x)\,\widehat{m}^c_{\mathtt{m}}(x) &= \epsilon^{abc}\lim_{2\mathfrak{a}\to0}\left[\widehat{J}^b_{L,\mathtt{m}}(x+2\mathfrak{a})+\widehat{J}^b_{R,\mathtt{m}}(x+2\mathfrak{a})\right]\left[\widehat{J}^c_{L,\mathtt{m}}(x)+\widehat{J}^c_{R,\mathtt{m}}(x)\right] \\ &= \epsilon^{abc}\lim_{2\mathfrak{a}\to0}\left[\widehat{J}^b_{L,\mathtt{m}}(x+2\mathfrak{a})\widehat{J}^c_{L,\mathtt{m}}(x)+\widehat{J}^b_{R,\mathtt{m}}(x+2\mathfrak{a})\widehat{J}^c_{R,\mathtt{m}}(x)+\widehat{J}^b_{L,\mathtt{m}}(x+2\mathfrak{a})\widehat{J}^c_{R,\mathtt{m}}(x)+\widehat{J}^b_{R,\mathtt{m}}(x+2\mathfrak{a})\widehat{J}^c_{L,\mathtt{m}}(x)\right] \\ &\sim \epsilon^{abc}\lim_{2\mathfrak{a}\to0}\left[\frac{\delta^{bc}}{8\pi^2(+\mathrm{i}2\mathfrak{a})^2}+\frac{\mathrm{i}\epsilon^{bcd}\,\widehat{J}^d_{L,\mathtt{m}}(x)}{2\pi(+\mathrm{i}2\mathfrak{a})}+\frac{\delta^{bc}}{8\pi^2(-\mathrm{i}2\mathfrak{a})^2}+\frac{\mathrm{i}\epsilon^{bcd}\,\widehat{J}^d_{R,\mathtt{m}}(x)}{2\pi(-\mathrm{i}2\mathfrak{a})}+0+0\right] \\ &= \lim_{2\mathfrak{a}\to0}\left[\frac{\mathrm{i}\epsilon^{abc}\epsilon^{bcd}\,\widehat{J}^d_{L,\mathtt{m}}(x)}{2\pi(+\mathrm{i}2\mathfrak{a})}+\frac{\mathrm{i}\epsilon^{abc}\epsilon^{bcd}\,\widehat{J}^d_{R,\mathtt{m}}(x)}{2\pi(-\mathrm{i}2\mathfrak{a})}\right] \\ &= \frac{(2\mathfrak{a})^{-1}}{\pi}\left[\widehat{J}^a_{L,\mathtt{m}}(x)-\widehat{J}^a_{R,\mathtt{m}}(x)\right].\end{aligned}\tag{A5a}$$

The OPE (2.18a) were used for the line with the $\sim$, the identity $\epsilon^{abc}\epsilon^{bcd} = 2\,\delta^{ad}$ was used to reach the last equality. Second,

$$\begin{aligned}\epsilon^{abc}\,\widehat{n}^{b}_{\mathtt{m}}(x)\,\widehat{n}^{c}_{\mathtt{m}}(x) &= \epsilon^{abc}\lim_{2\mathfrak{a}\to 0}\widehat{n}^{b}_{\mathtt{m}}(x+2\mathfrak{a})\,\widehat{n}^{c}_{\mathtt{m}}(x)\\ &\sim \epsilon^{abc}\,\frac{1}{2\pi^2\mathfrak{a}}\lim_{2\mathfrak{a}\to 0}\frac{\delta^{ab}}{|2\mathfrak{a}|}\\ &= 0.\end{aligned} \tag{A5b}$$

Hereto, the OPE (2.18b) were used. Third,

$$\begin{aligned}\epsilon^{abc}\,\widehat{m}^{b}_{\mathtt{m}}(x)\,\widehat{n}^{c}_{\mathtt{m}}(x) &= \epsilon^{abc}\lim_{2\mathfrak{a}\to 0}\Big[\widehat{J}^{b}_{L,\mathtt{m}}(x+2\mathfrak{a}) + \widehat{J}^{b}_{R,\mathtt{m}}(x+2\mathfrak{a})\Big]\,\widehat{n}^{c}_{\mathtt{m}}(x)\\ &= \epsilon^{abc}\lim_{2\mathfrak{a}\to 0}\Big[\widehat{J}^{b}_{L,\mathtt{m}}(x+2\mathfrak{a})\,\widehat{n}^{c}_{\mathtt{m}}(x) + \widehat{J}^{b}_{R,\mathtt{m}}(x+2\mathfrak{a})\,\widehat{n}^{c}_{\mathtt{m}}(x)\Big]\\ &\sim \epsilon^{abc}\lim_{2\mathfrak{a}\to 0}\left[\frac{\mathrm{i}\epsilon^{bcd}\,\widehat{n}^{d}_{\mathtt{m}}(x) + \mathrm{i}\delta^{bc}\,\widehat{\varepsilon}_{\mathtt{m}}(x)}{4\pi(+\mathrm{i}2\mathfrak{a})} + \frac{\mathrm{i}\epsilon^{bcd}\,\widehat{n}^{d}_{\mathtt{m}}(x) - \mathrm{i}\delta^{bc}\,\widehat{\varepsilon}_{\mathtt{m}}(x)}{4\pi(-\mathrm{i}2\mathfrak{a})}\right]\\ &= \lim_{2\mathfrak{a}\to 0}\left[\frac{\mathrm{i}\epsilon^{abc}\epsilon^{bcd}\,\widehat{n}^{d}_{\mathtt{m}}(x)}{4\pi(+\mathrm{i}2\mathfrak{a})} + \frac{\mathrm{i}\epsilon^{abc}\epsilon^{bcd}\,\widehat{n}^{d}_{\mathtt{m}}(x)}{4\pi(-\mathrm{i}2\mathfrak{a})}\right]\\ &= 0.\end{aligned} \tag{A5c}$$

The cancellation results here from two additive contributions of opposite chirality. Insertion of Eqs. (A5a) into Hamiltonian (A4) gives

$$\begin{aligned}\widehat{H}_{\triangle_1} \to\ &\chi_{\perp}\sum_{\mathtt{m}=1}^{n-1}\int_0^L \mathrm{d}x\,\frac{(2\mathfrak{a})}{\pi}\Big[\widehat{\boldsymbol{J}}_{L,\mathtt{m}+1}(x) + \widehat{\boldsymbol{J}}_{R,\mathtt{m}+1}(x)\Big]\cdot\Big[\widehat{\boldsymbol{J}}_{L,\mathtt{m}}(x) - \widehat{\boldsymbol{J}}_{R,\mathtt{m}}(x)\Big]\\ &+ \underline{\frac{\chi_{\perp}}{2}\sum_{\mathtt{m}=1}^{n-1}\int_0^L \mathrm{d}x\,\epsilon^{abc}\,(2\mathfrak{a})^3\left[\widehat{m}^{a}_{\mathtt{m}+1}(x) - \widehat{n}^{a}_{\mathtt{m}+1}(x)\right]\left[\partial_x\widehat{m}^{b}_{\mathtt{m}}(x) + \partial_x\widehat{n}^{b}_{\mathtt{m}}(x)\right]\left[\widehat{m}^{c}_{\mathtt{m}}(x) - \widehat{n}^{c}_{\mathtt{m}}(x)\right].}\end{aligned} \tag{A6}$$

Next, we write

$$\begin{aligned}\widehat{H}_{\triangle_2} :=\ &-\frac{\chi_{\perp}}{2}\sum_{i=1}^{N}\sum_{\mathtt{m}=1}^{n-1}\epsilon^{abc}\widehat{S}^{a}_{i+1,\mathtt{m}}\,\widehat{S}^{b}_{i+1,\mathtt{m}+1}\,\widehat{S}^{c}_{i,\mathtt{m}+1}\\ =\ &-\frac{\chi_{\perp}}{2}\sum_{i=1}^{N/2}\sum_{\mathtt{m}=1}^{n-1}\epsilon^{abc}\left(\widehat{S}^{a}_{2i+1,\mathtt{m}}\,\widehat{S}^{b}_{2i+1,\mathtt{m}+1}\,\widehat{S}^{c}_{2i,\mathtt{m}+1}\right)\\ &-\frac{\chi_{\perp}}{2}\sum_{i=1}^{N/2}\sum_{\mathtt{m}=1}^{n-1}\epsilon^{abc}\left(\widehat{S}^{a}_{2i,\mathtt{m}}\,\widehat{S}^{b}_{2i,\mathtt{m}+1}\,\widehat{S}^{c}_{2i-1,\mathtt{m}+1}\right),\end{aligned} \tag{A7}$$

where we have assumed that $N$ is an even integer and imposed PBC. If we insert the decomposition (3.8b) into the Hamiltonian (A7), we get

$$\begin{aligned}\widehat{H}_{\triangle_2} \to &-\frac{\chi_{\perp}}{2}\sum_{\mathtt{m}=1}^{n-1}\int_0^L \mathrm{d}x\,\epsilon^{abc}(2\mathfrak{a})^2\left[\widehat{m}^{a}_{\mathtt{m}}(x) - \widehat{n}^{a}_{\mathtt{m}}(x)\right]\left[\widehat{m}^{b}_{\mathtt{m}+1}(x) - \widehat{n}^{b}_{\mathtt{m}+1}(x)\right]\left[\widehat{m}^{c}_{\mathtt{m}+1}(x) + \widehat{n}^{c}_{\mathtt{m}+1}(x)\right]\\ &-\frac{\chi_{\perp}}{2}\sum_{\mathtt{m}=1}^{n-1}\int_0^L \mathrm{d}x\,\epsilon^{abc}(2\mathfrak{a})^2\left[\widehat{m}^{a}_{\mathtt{m}}(x) + \widehat{n}^{a}_{\mathtt{m}}(x)\right]\left[\widehat{m}^{b}_{\mathtt{m}+1}(x) + \widehat{n}^{b}_{\mathtt{m}+1}(x)\right]\underline{\left[\widehat{m}^{c}_{\mathtt{m}+1}(x-2\mathfrak{a}) - \widehat{n}^{c}_{\mathtt{m}+1}(x-2\mathfrak{a})\right]}.\end{aligned} \tag{A8}$$

To leading order in an expansion in powers of $(2\mathfrak{a})$,

$$\begin{aligned}
\widehat{H}_{\triangle_2} \to &-\frac{\chi_\perp}{2}\sum_{\mathtt{m}=1}^{n-1}\int_0^L \mathrm{d}x\,\epsilon^{abc}(2\mathfrak{a})^{1/2}\left[\widehat{m}^a_{\mathtt{m}}(x)-\widehat{n}^a_{\mathtt{m}}(x)\right]\left[\widehat{m}^b_{\mathtt{m}+1}(x)-\widehat{n}^b_{\mathtt{m}+1}(x)\right]\left[\widehat{m}^c_{\mathtt{m}+1}(x)+\widehat{n}^c_{\mathtt{m}+1}(x)\right]\\
&-\frac{\chi_\perp}{2}\sum_{\mathtt{m}=1}^{n-1}\int_0^L \mathrm{d}x\,\epsilon^{abc}(2\mathfrak{a})^{2}\left[\widehat{m}^a_{\mathtt{m}}(x)+\widehat{n}^a_{\mathtt{m}}(x)\right]\left[\widehat{m}^b_{\mathtt{m}+1}(x)+\widehat{n}^b_{\mathtt{m}+1}(x)\right]\underline{\left[\widehat{m}^c_{\mathtt{m}+1}(x)-\widehat{n}^c_{\mathtt{m}+1}(x)\right]}\\
&+\frac{\chi_\perp}{2}\sum_{\mathtt{m}=1}^{n-1}\int_0^L \mathrm{d}x\,\epsilon^{abc}(2\mathfrak{a})^{3}\left[\widehat{m}^a_{\mathtt{m}}(x)+\widehat{n}^a_{\mathtt{m}}(x)\right]\left[\widehat{m}^b_{\mathtt{m}+1}(x)+\widehat{n}^b_{\mathtt{m}+1}(x)\right]\underline{\left[\partial_x\widehat{m}^c_{\mathtt{m}+1}(x)-\partial_x\widehat{n}^c_{\mathtt{m}+1}(x)\right]}.
\end{aligned}\tag{A9}$$

where the first and the second line can be further simplified,

$$\begin{aligned}
\widehat{H}_{\triangle_2} \to &-\frac{\chi_\perp}{2}\sum_{\mathtt{m}=1}^{n-1}\int_0^L \mathrm{d}x\,\epsilon^{abc}\,2\times(2\mathfrak{a})^{2}\Big[\widehat{m}^a_{\mathtt{m}}(x)\,\widehat{m}^b_{\mathtt{m}+1}(x)\,\widehat{m}^c_{\mathtt{m}+1}(x)-\widehat{m}^a_{\mathtt{m}}(x)\,\widehat{n}^b_{\mathtt{m}+1}(x)\,\widehat{n}^c_{\mathtt{m}+1}(x)\\
&\qquad\qquad -\widehat{n}^a_{\mathtt{m}}(x)\,\widehat{m}^b_{\mathtt{m}+1}(x)\,\widehat{n}^c_{\mathtt{m}+1}(x)+\widehat{n}^a_{\mathtt{m}}(x)\,\widehat{n}^b_{\mathtt{m}+1}(x)\,\widehat{m}^c_{\mathtt{m}+1}(x)\Big]\\
&+\frac{\chi_\perp}{2}\sum_{\mathtt{m}=1}^{n-1}\int_0^L \mathrm{d}x\,\epsilon^{abc}(2\mathfrak{a})^{3}\left[\widehat{m}^a_{\mathtt{m}}(x)+\widehat{n}^a_{\mathtt{m}}(x)\right]\left[\widehat{m}^b_{\mathtt{m}+1}(x)+\widehat{n}^b_{\mathtt{m}+1}(x)\right]\left[\partial_x\widehat{m}^c_{\mathtt{m}+1}(x)-\partial_x\widehat{n}^c_{\mathtt{m}+1}(x)\right].
\end{aligned}\tag{A10}$$

To proceed, we need to point-splitting pairs of operators sitting at the same position $x$ in the same bundle $\mathtt{m}$. After point-splitting, we shall use the OPE (2.18) so as to reduce the point-split pair of operators to either a $\mathbb{C}$ number or a single operator. Insertion of Eqs. (A5a) into Hamiltonian (A10) gives

$$\begin{aligned}
\widehat{H}_{\triangle_2} \to &-\chi_\perp\sum_{\mathtt{m}=1}^{n-1}\int_0^L \mathrm{d}x\,\frac{(2\mathfrak{a})}{\pi}\Big[\widehat{\boldsymbol{J}}_{L,\mathtt{m}}(x)+\widehat{\boldsymbol{J}}_{R,\mathtt{m}}(x)\Big]\cdot\Big[\widehat{\boldsymbol{J}}_{L,\mathtt{m}+1}(x)-\widehat{\boldsymbol{J}}_{R,\mathtt{m}+1}(x)\Big]\\
&\underline{+\frac{\chi_\perp}{2}\sum_{\mathtt{m}=1}^{n-1}\int_0^L \mathrm{d}x\,\epsilon^{abc}(2\mathfrak{a})^{3}\left[\widehat{m}^a_{\mathtt{m}}(x)+\widehat{n}^a_{\mathtt{m}}(x)\right]\left[\widehat{m}^b_{\mathtt{m}+1}(x)+\widehat{n}^b_{\mathtt{m}+1}(x)\right]\left[\partial_x\widehat{m}^c_{\mathtt{m}+1}(x)-\partial_x\widehat{n}^c_{\mathtt{m}+1}(x)\right].}
\end{aligned}\tag{A11}$$

Observe that the second lines (the underlined ones) in Eqs. (A6) and (A11) are irrelevant perturbations.

Thus, the continuum limit of $\widehat{H}_{\triangle}$ (4.1a) can be read from Eqs. (A6) and (A11). It is

$$\begin{aligned}
\widehat{H}_{\triangle} =&(2\mathfrak{a})\frac{\chi_\perp}{\pi}\sum_{\mathtt{m}=1}^{n-1}\int_0^L \mathrm{d}x\,\Big[\Big(\widehat{\boldsymbol{J}}_{L,\mathtt{m}}-\widehat{\boldsymbol{J}}_{R,\mathtt{m}}\Big)\cdot\Big(\widehat{\boldsymbol{J}}_{L,\mathtt{m}+1}+\widehat{\boldsymbol{J}}_{R,\mathtt{m}+1}\Big)-\Big(\widehat{\boldsymbol{J}}_{L,\mathtt{m}}+\widehat{\boldsymbol{J}}_{R,\mathtt{m}}\Big)\cdot\Big(\widehat{\boldsymbol{J}}_{L,\mathtt{m}+1}-\widehat{\boldsymbol{J}}_{R,\mathtt{m}+1}\Big)\Big]+\cdots\\
=&g_{\chi_\perp}\sum_{\mathtt{m}=1}^{n-1}\int_0^L \mathrm{d}x\,\Big(\widehat{\boldsymbol{J}}_{L,\mathtt{m}}\cdot\widehat{\boldsymbol{J}}_{R,\mathtt{m}+1}-\widehat{\boldsymbol{J}}_{R,\mathtt{m}}\cdot\widehat{\boldsymbol{J}}_{L,\mathtt{m}+1}\Big)+\cdots,
\end{aligned}\tag{A12a}$$

where $\cdots$ refers to irrelevant local perturbations and where the coupling $g_{\chi_\perp}$ stands for

$$g_{\chi_\perp}=2\times(2\mathfrak{a})\,\frac{\chi_\perp}{\pi}.\tag{A12b}$$

Similarly, the continuum limit of $\widehat{H}'_{\triangle}$ follows from that of $\widehat{H}_{\triangle}$ with the substitution $\boldsymbol{J}_{M,\mathtt{m}}(x)\to\boldsymbol{J}'_{M,\mathtt{m}}(x)$ with $M=L,R$.

### Appendix B: The nonvanishing bare couplings (4.4b–4.4e)

We seek the naive continuum limit of $\widehat{H}_{\square}$ defined by Eq. (4.1b). To this end, we first isolate from $\widehat{H}_{\square}$ the contribution

$$\widehat{H}_{J_\perp}:=J_\perp\sum_{i=1}^{N/2}\sum_{\mathtt{m}=1}^{n-1}\widehat{\boldsymbol{S}}_{2i,\mathtt{m}}\cdot\widehat{\boldsymbol{S}}_{2i,\mathtt{m}+1}+J_\perp\sum_{i=1}^{N/2}\sum_{\mathtt{m}=1}^{n-1}\widehat{\boldsymbol{S}}_{2i+1,\mathtt{m}}\cdot\widehat{\boldsymbol{S}}_{2i+1,\mathtt{m}+1},\tag{B1}$$

where we have assumed that $N$ is an even integer and imposed PBC. If we insert the decomposition (3.8) into the chain-like Hamiltonian (B1), we get

$$\begin{aligned}\widehat{H}_{J_\perp} &\to (2\mathfrak{a})J_\perp \sum_{\mathtt{m}=1}^{n-1}\int_0^L \mathrm{d}x\,\left[\widehat{\boldsymbol{m}}_{\mathtt{m}}(x)+\widehat{\boldsymbol{n}}_{\mathtt{m}}(x)\right]\cdot\left[\widehat{\boldsymbol{m}}_{\mathtt{m}+1}(x)+\widehat{\boldsymbol{n}}_{\mathtt{m}+1}(x)\right]\\ &\quad+(2\mathfrak{a})J_\perp \sum_{\mathtt{m}=1}^{n-1}\int_0^L \mathrm{d}x\,\left[\widehat{\boldsymbol{m}}_{\mathtt{m}}(x)-\widehat{\boldsymbol{n}}_{\mathtt{m}}(x)\right]\cdot\left[\widehat{\boldsymbol{m}}_{\mathtt{m}+1}(x)-\widehat{\boldsymbol{n}}_{\mathtt{m}+1}(x)\right]\\ &=2\times(2\mathfrak{a})J_\perp \sum_{\mathtt{m}=1}^{n-1}\int_0^L \mathrm{d}x\,\left[\widehat{\boldsymbol{m}}_{\mathtt{m}}(x)\cdot\widehat{\boldsymbol{m}}_{\mathtt{m}+1}(x)+\widehat{\boldsymbol{n}}_{\mathtt{m}}(x)\cdot\widehat{\boldsymbol{n}}_{\mathtt{m}+1}(x)\right].\end{aligned}\tag{B2}$$

No gradient expansion was needed in this step. Hence, the quantum critical point at $J_2=J_\vee=J_\perp=J_\backslash=J_/=0$ acquires the local perturbation

$$\widehat{V}^{\perp}_{\mathrm{per}}(x) := \sum_{\mathtt{m}=1}^{n-1}\left(\widehat{W}^{\perp}_{jj,\mathtt{m}}+\widehat{W}^{\perp}_{nn,\mathtt{m}}\right)(x),\tag{B3a}$$

where

$$\widehat{W}^{\perp}_{jj,\mathtt{m}}(x) := g^{\perp}_{jj}\left[\widehat{\boldsymbol{J}}_{L,\mathtt{m}}(x)+\widehat{\boldsymbol{J}}_{R,\mathtt{m}}(x)\right]\cdot\left[\widehat{\boldsymbol{J}}_{L,\mathtt{m}+1}(x)+\widehat{\boldsymbol{J}}_{R,\mathtt{m}+1}(x)\right],\qquad g^{\perp}_{jj} := 2\times v_\perp,\qquad v_\perp := (2\mathfrak{a})\,J_\perp,\tag{B3b}$$

$$\widehat{W}^{\perp}_{nn,\mathtt{m}}(x) := g^{\perp}_{nn}\,\widehat{\boldsymbol{n}}_{\mathtt{m}}(x)\cdot\widehat{\boldsymbol{n}}_{\mathtt{m}+1}(x),\qquad g^{\perp}_{nn} := 2\times v_\perp.\tag{B3c}$$

Second, we isolate from $\widehat{H}_\square$ defined by Eq. (4.1b) the contribution

$$\widehat{H}_{J_/} := J_/ \sum_{i=1}^{N/2}\sum_{\mathtt{m}=1}^{n-1}\widehat{\boldsymbol{S}}_{2i,\mathtt{m}}\cdot\widehat{\boldsymbol{S}}_{2i+1,\mathtt{m}+1}+J_/ \sum_{i=1}^{N/2}\sum_{\mathtt{m}=1}^{n-1}\widehat{\boldsymbol{S}}_{2i+1,\mathtt{m}}\cdot\widehat{\boldsymbol{S}}_{2i+2,\mathtt{m}+1},\tag{B4}$$

where we have assumed that $N$ is an even integer and imposed PBC. If we insert the decomposition (3.8) into the chain-like Hamiltonian (B4), we get

$$\begin{aligned}\widehat{H}_{J_/} &\to (2\mathfrak{a})\,J_/ \sum_{\mathtt{m}=1}^{n-1}\int_0^L \mathrm{d}x\,\left[\widehat{\boldsymbol{m}}_{\mathtt{m}}(x)+\widehat{\boldsymbol{n}}_{\mathtt{m}}(x)\right]\cdot\left[\widehat{\boldsymbol{m}}_{\mathtt{m}+1}(x)-\widehat{\boldsymbol{n}}_{\mathtt{m}+1}(x)\right]\\ &\quad+(2\mathfrak{a})\,J_/ \sum_{\mathtt{m}=1}^{n-1}\int_0^L \mathrm{d}x\,\left[\widehat{\boldsymbol{m}}_{\mathtt{m}}(x)-\widehat{\boldsymbol{n}}_{\mathtt{m}}(x)\right]\cdot\left[\widehat{\boldsymbol{m}}_{\mathtt{m}+1}(x+2\mathfrak{a})+\widehat{\boldsymbol{n}}_{\mathtt{m}+1}(x+2\mathfrak{a})\right].\end{aligned}\tag{B5}$$

To leading order in an expansion in powers of $(2\mathfrak{a})$,

$$\begin{aligned}\widehat{H}_{J_/} &\to (2\mathfrak{a})\,J_/ \sum_{\mathtt{m}=1}^{n-1}\int_0^L \mathrm{d}x\,\left[\widehat{\boldsymbol{m}}_{\mathtt{m}}(x)+\widehat{\boldsymbol{n}}_{\mathtt{m}}(x)\right]\cdot\left[\widehat{\boldsymbol{m}}_{\mathtt{m}+1}(x)-\widehat{\boldsymbol{n}}_{\mathtt{m}+1}(x)\right]\\ &\quad+(2\mathfrak{a})\,J_/ \sum_{\mathtt{m}=1}^{n-1}\int_0^L \mathrm{d}x\,\left[\widehat{\boldsymbol{m}}_{\mathtt{m}}(x)-\widehat{\boldsymbol{n}}_{\mathtt{m}}(x)\right]\cdot\left[\widehat{\boldsymbol{m}}_{\mathtt{m}+1}(x)+\widehat{\boldsymbol{n}}_{\mathtt{m}+1}(x)+(2\mathfrak{a})\partial_x\widehat{\boldsymbol{m}}_{\mathtt{m}+1}(x)+(2\mathfrak{a})\partial_x\widehat{\boldsymbol{n}}_{\mathtt{m}+1}(x)+\mathcal{O}\big((2\mathfrak{a})^2\big)\right]\\ &=(2\mathfrak{a})J_/ \sum_{\mathtt{m}=1}^{n-1}\int_0^L \mathrm{d}x\,\left[2\times\widehat{\boldsymbol{m}}_{\mathtt{m}}(x)\cdot\widehat{\boldsymbol{m}}_{\mathtt{m}+1}(x)-2\times\widehat{\boldsymbol{n}}_{\mathtt{m}}(x)\cdot\widehat{\boldsymbol{n}}_{\mathtt{m}+1}(x)-(2\mathfrak{a})\,\widehat{\boldsymbol{n}}_{\mathtt{m}}(x)\cdot\partial_x\widehat{\boldsymbol{n}}_{\mathtt{m}+1}(x)+\cdots\right],\end{aligned}\tag{B6}$$

where $\cdots$ denotes irrelevant local perturbations. Hence, the quantum critical point at $J_2=J_\vee=J_\perp=J_\backslash=J_/=0$ acquires the local perturbation

$$\widehat{V}^{/}_{\mathrm{per}}(x) := \sum_{\mathtt{m}=1}^{n-1}\left(\widehat{W}^{/}_{jj,\mathtt{m}}+\widehat{W}^{/}_{nn,\mathtt{m}}+\widehat{W}^{/}_{\mathrm{tw},\mathtt{m}}\right)(x),\tag{B7a}$$

where

$$\widehat{W}^{/}_{jj,\mathtt{m}}(x) \coloneqq g^{/}_{jj}\,\left[\widehat{\boldsymbol{J}}_{L,\mathtt{m}}(x)+\widehat{\boldsymbol{J}}_{R,\mathtt{m}}(x)\right]\cdot\left[\widehat{\boldsymbol{J}}_{L,\mathtt{m}+1}(x)+\widehat{\boldsymbol{J}}_{R,\mathtt{m}+1}(x)\right], \qquad g^{/}_{jj} \coloneqq 2\times v_{/}, \qquad v_{/} \coloneqq (2\mathfrak{a})\,J_{/}, \tag{B7b}$$

$$\widehat{W}^{/}_{nn,\mathtt{m}}(x) \coloneqq g^{/}_{nn}\,\widehat{\boldsymbol{n}}_{\mathtt{m}}(x)\cdot\widehat{\boldsymbol{n}}_{\mathtt{m}+1}(x), \qquad g^{/}_{nn} \coloneqq -2\times v_{/}, \tag{B7c}$$

$$\widehat{W}^{/}_{\mathrm{tw},\mathtt{m}}(x) \coloneqq g^{/}_{\mathrm{tw}}\,\widehat{\boldsymbol{n}}_{\mathtt{m}}(x)\cdot\partial_x\widehat{\boldsymbol{n}}_{\mathtt{m}+1}(x), \qquad g^{/}_{\mathrm{tw}} \coloneqq -(2\mathfrak{a})\,v_{/}. \tag{B7d}$$

Third and at last, we isolate from $\widehat{H}_{\square}$ defined by Eq. (4.1b) the contribution

$$\widehat{H}_{J_{\backslash}} \coloneqq J_{\backslash}\sum_{i=1}^{N/2}\sum_{\mathtt{m}=1}^{n-1}\widehat{\boldsymbol{S}}_{2i,\mathtt{m}+1}\cdot\widehat{\boldsymbol{S}}_{2i+1,\mathtt{m}} + J_{\backslash}\sum_{i=1}^{N/2}\sum_{\mathtt{m}=1}^{n-1}\widehat{\boldsymbol{S}}_{2i+1,\mathtt{m}+1}\cdot\widehat{\boldsymbol{S}}_{2i+2,\mathtt{m}}, \tag{B8}$$

where we have assumed that $N$ is an even integer and imposed PBC. Observe that the continuum limit of Eq. (B8) can be obtained from the continuum limit of Eq. (B4) with the interchange of $\mathtt{m}$ and $\mathtt{m}+1$ in Eq. (B5). Hence, the quantum critical point at $J_2 = J_{\vee} = J_{\perp} = J_{\backslash} = J_{/} = 0$ acquires the local perturbation

$$\widehat{V}^{\backslash}_{\mathrm{per}}(x) \coloneqq \sum_{\mathtt{m}=1}^{n-1}\left(\widehat{W}^{\backslash}_{jj,\mathtt{m}}+\widehat{W}^{\backslash}_{nn,\mathtt{m}}+\widehat{W}^{\backslash}_{\mathrm{tw},\mathtt{m}}\right)(x), \tag{B9a}$$

where

$$\widehat{W}^{\backslash}_{jj,\mathtt{m}}(x) \coloneqq g^{\backslash}_{jj}\,\left[\widehat{\boldsymbol{J}}_{L,\mathtt{m}+1}(x)+\widehat{\boldsymbol{J}}_{R,\mathtt{m}+1}(x)\right]\cdot\left[\widehat{\boldsymbol{J}}_{L,\mathtt{m}}(x)+\widehat{\boldsymbol{J}}_{R,\mathtt{m}}(x)\right], \qquad g^{\backslash}_{jj} \coloneqq 2\times v_{\backslash}, \qquad v_{\backslash} \coloneqq (2\mathfrak{a})\,J_{\backslash}, \tag{B9b}$$

$$\widehat{W}^{\backslash}_{nn,\mathtt{m}}(x) \coloneqq g^{\backslash}_{nn}\,\widehat{\boldsymbol{n}}_{\mathtt{m}+1}(x)\cdot\widehat{\boldsymbol{n}}_{\mathtt{m}}(x), \qquad g^{\backslash}_{nn} \coloneqq -2\times v_{\backslash}, \tag{B9c}$$

$$\widehat{W}^{\backslash}_{\mathrm{tw},\mathtt{m}}(x) \coloneqq g^{\backslash}_{\mathrm{tw}}\,\widehat{\boldsymbol{n}}_{\mathtt{m}+1}(x)\cdot\partial_x\widehat{\boldsymbol{n}}_{\mathtt{m}}(x) = -g^{\backslash}_{\mathrm{tw}}\,\widehat{\boldsymbol{n}}_{\mathtt{m}}(x)\cdot\partial_x\widehat{\boldsymbol{n}}_{\mathtt{m}+1}(x) + \text{total derivative}, \qquad g^{\backslash}_{\mathrm{tw}} \coloneqq -(2\mathfrak{a})\,v_{\backslash}. \tag{B9d}$$

[1] H. Sugawara, Phys. Rev. **170**, 1659 (1968).
[2] O. A. Starykh, A. Furusaki, and L. Balents, Phys. Rev. B **72**, 094416 (2005).
[3] S. Furukawa, M. Sato, S. Onoda, and A. Furusaki, Phys. Rev. B **86**, 094417 (2012).
[4] A. Metavitsiadis, D. Sellmann, and S. Eggert, Phys. Rev. B **89**, 241104 (2014).
[5] F. D. M. Haldane, Phys. Rev. B **25**, 4925 (1982).
[6] C. K. Majumdar and D. K. Ghosh, J. Math. Phys. **10**, 1399 (1969).
[7] C. Mudry and E. Fradkin, Phys. Rev. B **50**, 11409 (1994).
[8] P.-H. Huang, J.-H. Chen, P. R. S. Gomes, T. Neupert, C. Chamon, and C. Mudry, Phys. Rev. B **93**, 205123 (2016).
[9] A. A. Nersesyan, A. O. Gogolin, and F. H. L. Eßler, Phys. Rev. Lett. **81**, 910 (1998).
[10] D. Allen, F. H. L. Essler, and A. A. Nersesyan, Phys. Rev. B **61**, 8871 (2000).
[11] O. A. Starykh and L. Balents, Phys. Rev. Lett. **98**, 077205 (2007).
[12] G. Gorohovsky, R. G. Pereira, and E. Sela, Phys. Rev. B **91**, 245139 (2015).